\documentclass{article}

\usepackage{arxiv}
\RequirePackage{amsthm,amssymb,amsmath,longtable,color, colortbl, float, nicematrix, mathtools}

\usepackage[utf8]{inputenc} 
\usepackage[T1]{fontenc}    
\usepackage{hyperref}       
\usepackage{url}            
\usepackage{booktabs}       
\usepackage{amsfonts}       
\usepackage{nicefrac}       
\usepackage{microtype}      
\usepackage{cleveref}       
\usepackage{lipsum}         
\usepackage{graphicx}
\usepackage{natbib}
\usepackage{doi}

\mathtoolsset{showonlyrefs=true}
   \usepackage{natbib}
    \bibpunct{(}{)}{;}{a}{,}{,}
    \usepackage{xpatch}
    \xpatchbibdriver{article}
    {\printfield{year}\newunit}
    {\printfield{year}}
    {}{}
\title{Spatio-temporal analysis of extreme winter temperatures in Ireland}

\usepackage{authblk}

\author{\large{Dáire Healy$^{1*}$, Jonathan A. Tawn$^{2}$, Peter Thorne$^{3}$ and Andrew Parnell$^{1}$}
}
\normalsize
\vspace{1em}
\affil{$^1$School of Mathematics and Statistics, University College Dublin, Dublin, Ireland}

\affil{$^2$Department of Mathematics and Statistics, Lancaster University, Lancaster, UK}

\affil{$^2$Irish Climate Analysis and Research UnitS (ICARUS), Department of
Geography, Maynooth University, Maynooth, Ireland}

\affil{$^*$\texttt{\url{daire.healy@ucd.ie}}}

\date{}
\begin{document}
\maketitle
\begin{abstract}
We analyse synoptic observational data on extreme daily minimum temperatures in winter months over Ireland from 1950--2022. We model the tail of the marginal distributions of extreme winter minima using a generalised Pareto distribution capturing temporal and spatial sources of non-stationarity.  We disentangle long-term climate trends from obfuscating shorter-term (month-to-winter long) large fluctuations, caused by e.g., anomalous behaviour of the jet stream.  We identify extreme spatial events with respect to a carefully chosen risk function and fit an $r$-Pareto process to extreme events exceeding a high-risk threshold. We show that long-term trends over Ireland of extremely cold winter temperatures are warming at a faster rate than both mean winter temperatures and summer extreme daily maximum temperatures. Critically, we show that if we did not account for large-scale, short-term climatic oscillations, we would incorrectly estimate that the extremely cold winter temperatures were getting colder. In terms of spatial extreme events, we find that in periods where the jet stream typically produces the coldest winter temperatures, the estimated rate at which temperatures could fall below the coldest value ever recorded has decreased by a factor of 100 over the study period
\end{abstract}


\section{Introduction}\label{sec:c6_intro}
The frequency and intensity of extremely cold temperatures have decreased globally since 1950, with this trend expected to continue as global mean temperatures rise \citep[Chapter 11]{IPCC2021}. Winter temperatures have been particularly warming in the northern mid-latitudes (\citealt{Heidrun2015}, \citealt{VanOldenborgh2019}). In Ireland, significant warming of minimum air temperatures has been observed, with a decreasing frequency of cold events since 1960 over different seasons 
\citep{McElwain2003}, which is in line with global trends (\citealp{Nolan2020}, \citealp{Mateus2022}, \citealp{Garcia2022}). Globally, there is strong evidence that minimum temperatures are increasing faster than maxima  \citep{Huang2016, Rhines2017, dunn2020, Krock2022}. Extremely high temperatures receive a substantially disproportionate amount of analysis in the literature, given their direct and immediate link to loss of life \citep{Ballester2023}, crop failure \citep{He2022}, etc. Furthermore, the intensity, duration and extent of heatwaves are expected to increase with global temperatures \citep{Perkins2020}, exacerbating impacts and demanding immediate attention. While the warming of winter temperatures is less immediately disruptive, it nonetheless harbours many potentially devastating consequences. Changes in winter extremes are having an effect on many facets of our environment and society, from mortality and morbidity rates \citep{Conlon2011}, to agricultural activity \citep{bindi2011, hooker2008} and ecosystems \citep{Osland2021}. For example, many regions rely on cold winters to control pathogens and pest populations \citep{Skend2021}. It has been argued that the consequences of warming winters have been understudied \citep{Boucek2016}.

Ireland's winters are generally less severe compared to many countries at similar latitudes due to the regulating effect of the North Atlantic Ocean and the Irish Sea surrounding the island. The winter of 2019 was the warmest on record for Ireland at 0.9°C above the 1961--1990 average winter temperature \citep{Meteireann2019}. However, in northern Europe, warming trends are contrasted with occasional, extreme cold events associated with variability in climatic oscillations due to changes in the jet stream, i.e., Arctic Oscillation (AO) and the North Atlantic Oscillation (NAO) \citep{Vihma2020}. Exploratory investigations by \cite{McElwain2003} have identified significant relationships between temperature and the NAO index in winter on the west coast of Ireland. They note that `the future course of Irish climate [analysis] is undoubtedly going to be linked with large-scale atmospheric circulation". The coldest temperatures in Ireland since 1900 were observed as recently as 2010, e.g., a record low of -17.5°C was observed in Co.~Mayo, Straide \citep{CHRISTIANSEN2018}. The extremely cold winter of 2009/10 was caused by an anomalously persistent negative phase of the NAO \citep{Cattiaux2010}. A similar event occurred in the winter of 1962/63, where a Scandinavian anticyclone, marked by a negative phase of the NAO, brought easterly winds over Ireland, resulting in extremely cold temperatures \citep{Sippel2024}. Throughout the analysis, we highlight the winter of 2009/10 as the primary example of an unusually cold winter event, given its more recent occurrence and its more abundant data availability.

\begin{figure}[h]
    \centering
    \includegraphics[width = 12cm]{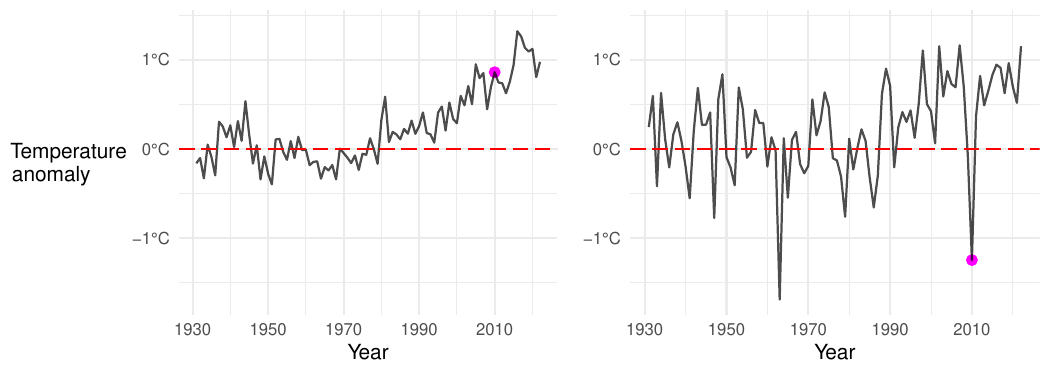}
    \caption[Global and Irish mean winter temperature anomalies with 2010 highlighted.]{Global (left) and Irish (right) mean winter (DJF) temperature anomalies with 2010 highlighted. Values are calculated from the HadCRUT5 data product. Anomalies are relative to a 1961-1990 reference period.}
    \label{fig:hadcrut_irel_v_glob}
\end{figure}

Figure~\ref{fig:hadcrut_irel_v_glob} shows that globally (left-hand plot), the winter months of December, January, and February (which we subsequently denote by DJF) of 2010 were anomalously warm, despite large variations locally, e.g., in Ireland (right-hand plot), with these values derived from the HadCRUT5 data product, see Section~\ref{sec:c6_short_term_covariates} for details. Disentangling these juxtaposing trends in winter extremes is important to fully ascertain the overall trend of winter extreme temperatures. \cite{CHRISTIANSEN2018} show that the cold winter of 2009/10 was, in fact, warmer than expected, in the context of the climate conditions at the time. 
Based on a range of different climate model ensemble runs
\cite{CHRISTIANSEN2018} estimate that the occurrence probability of extremely cold winter temperatures, such as those seen in 2009/10, has reduced by a factor of two due to anthropogenically induced climate change, and so were in line with global warming winter trends. However, their analysis does not examine how these climatic features influence the spatial patterns of extremely cold winter temperatures, nor does it assess whether such features are evident in observational data.

Our analysis aims to test for and identify non-stationarities in winter temperature extremes to give a more thorough picture of the impact of climate change on the island of Ireland (i.e., the Republic of Ireland and Northern Ireland), hereafter simply referred to as Ireland. We aim to characterise trends in winter temperatures in terms of extreme events at single sites but also in terms of the occurrence of extreme spatial events with specific characteristics, e.g., somewhere in Ireland is colder than $T^\circ C$. The analysis and methodology presented in this paper are complementary to the statistical methods and associated data analysis on extremely hot summer temperatures in Ireland performed by \cite{Healy2023}, and present formal strategies for assessing the effect of climate indices as motivated by the analysis of \cite{McElwain2003}. We emphasise novel aspects of our modelling strategy, such as our much deeper dive into using physically based covariates, as well as responding here to issues raised in the discussion contributions of \cite{Healy2023}. We provide new insights into the behaviour of extremely cold temperatures in Ireland and set out a framework for interpreting for how they are related to different behaviours of the jet stream. We ensure that the material presented here is entirely self-contained, but in places, we cross-reference to our previous work for more details. 
Here, our interest is in lower tail extreme events, and so by negating the data, we perform the analysis on upper tail events of these negated data. Then, by negating the resulting estimated quantiles, we report all our estimates on the scale of the original temperature data.

In common with \cite{Healy2023}, we aim to avoid bias in estimating the severity of extreme weather events by relying on asymptotically justified statistical models provided by extreme value theory. For marginal extreme value modelling, we take the standard approach of adopting a high threshold and fitting a generalised Pareto distribution (GPD) to excesses \citep{Coles2001}. As extreme spatial temperature events are not likely to exceed the marginal thresholds at all sites over Ireland simultaneously, we also need to model the marginal distributions below the thresholds. We blend together GPD and non-parametric quantile regression estimates for our hybrid estimator of the marginal distributions. We use spatio-temporal regression models for the GPD parameters and non-parametric elements, whilst we use only spatial regression models for the threshold function. Threshold estimation is well known to be a challenging task, and our focus is on the temporal changes in extreme events, so we do not want the subjective choice of the threshold inference to bias temporal findings.

{Advances in the literature on extreme value theory and methods have introduced a suite of models for characterising the spatial dependence of extreme observations. Since we want to model the behaviour of actual spatial extreme events, this restricts the set of appropriate statistical methodologies that are motivated by asymptotic theory. In particular, it rules out using max-stable process models, which are only asymptotically justified as models for site-wise maxima, which typically do not correspond to observed concurrent spatial events \citep{Huser2025}. Instead, we use the $r$-Pareto process, which is a flexible limiting spatial process conditioning on some feature of the process being extreme \citep{DeFondeville2018}. The $r$-Pareto process is a particular type of asymptotically motivated Pareto spatial process. Generalised Pareto processes were introduced by \cite{Buishand2008}, see also \cite{Ferreira2014}, and were constructed by conditioning on the supremum of the process being large. \cite{Dombry2015} proposed a relaxation of this conditioning event, by extending the framework to conditioning on any first-order homogenous functional (which includes the supremum) of the process being large, with the resulting limit process being termed an $\ell$-Pareto process, and later called an $r$-Pareto process \citep{DeFondeville2018}.  Since then, functional exceedances of extremal fields have been analysed using $r$-Pareto processes in several settings, partly for their flexibility but also due to the accessibility of publicly available software such as \texttt{mvpot} \citep{DeFondeville2021}. Our experience of the different choices of homogeneous functional is that the results of subsequent inference are largely insensitive to this choice; see our discussion in Section~\ref{sec:c6_rparp}. Pareto processes have been used in a range of spatial rainfall applications, for example, Mediterranean France \citep{Palacios2020} and Burkina Faso in West Africa \citep{Bewentaore2022}. We model the $r$-Pareto process on a transformed scale, after accounting for the different marginal features. As \cite{Koh2025} has suggested, it is possible to perform all the modelling on the temperature scale, as in \cite{DeFondeville2022}. Indeed, this would have been possible as our selected marginal model has a constant GPD shape parameter over space and time, but prior to modelling the data, we did not know that having such a simplification would be possible, so we took the route of analysis presented here.}

Spatial Pareto processes impose limitations on the characteristics of extremal dependence structures that can be captured. In particular, they give either asymptotic dependence or exact independence between the process for any pair of sites, with the former imposing that there is a non-zero probability of an event being large at all sites given that it is large at one, see Section~\ref{sec:ADvAI} for mathematical limit details. Thus, Pareto processes do not capture well the extremes of dependent processes which are not asymptotically dependent (termed asymptotically independent processes), such as Gaussian copula processes \citep{Wadsworth2012}. For asymptotically independent processes, the spatial scale of events reduces as the marginal magnitude of the event increases, with this property frequently observed in environmental applications \citep{Huser2025}. Consequently, if the process is asymptotically independent, using methods based on an assumption of asymptotic dependence will lead to the overestimation of the probability of jointly extreme events over space. When a process exhibits asymptotic independence, there are theoretical and practical advantages in using methods for conditional spatial extremes \citep{Wadsworth2019, Simpson2021b, Richards2021} or max-infinite divisible processes \citep{Huser2021, Bopp2021, HealyProsdocimi2025}. Whether asymptotic dependence or asymptotic independence provides a better description of extremal dependence in a given application depends on how the scale of the spatial domain of interest compares with the spatial dependence scale of the process. This choice, in practice, can be informed only by empirical evidence. For our application, Ireland is small relative to cold weather processes in winter, so it is reasonable to expect that an $r$-Pareto process will be suitable, but it is critical to check that this assumption is realistic, which is part of our exploratory analysis in Section~\ref{sec:ADvAI}.

{This paper offers a number of scientifically and technically important extensions to the work of \cite{Healy2023}. Firstly, there is a marked, non-cosmetic, seasonal contrast in the level of modelling that is required to achieve a well-fitting model for cold winter extremes. Although both hot summers and cold winters have smooth long-term trends, the differences between the changes over winters are substantial relative to over summers, as winter temperature extremes are strongly influenced by different physical drivers (e.g., jet stream anomalies, NAO, AO, and summaries of temperature anomalies) which manifest in the form of non-stationarity due to short-term (intra-year) climatic variability. Hence, this study involves analysis based on extensive model choice over sets of feasible covariates that capture this short-term variability in addition to the spatial and temporal covariates considered by \cite{Healy2023}. Our analysis underlines the importance of considering short-term climatic variability when modelling extreme weather processes, for which we propose a novel framework. Secondly, to account for these short-term climatic variability, we have developed a customised spatio-temporal bootstrapping procedure which also addresses the strong temporal autocorrelation in cold temperature extremes and accounts for the extensive missing data, which can reasonably be viewed as missing at random \citep{little2019statistical}.}

Our results provide important and transparent scientific insight. In particular, our paper highlights that cold extremes are warming more rapidly than mean winter temperatures, and that winters are warming faster than hot summers. We find that the phase of any short-term climatic variability covariate is as important to winter cold temperatures as long-term trends over our study window, and that ignoring such a covariate would have led to a false conclusion that cold winter temperatures were cooling. Thus emphasising how important it is to incorporate such a covariate into the analysis. Inferences have been made on the rate of change of the spatial properties of extremely cold winter events. For example, we estimate that the rate of some stations in Ireland experiencing a colder temperature than $-19.4^\circ$C, the coldest ever recorded temperature in Ireland, has decreased by a factor of 100 from 1950 to 2022. Thus, we have found vital differences over time in the occurrence of extreme spatial events. 

Our paper is organised as follows. Section~\ref{sec:c6_data} details the observational and climate model data used, as well as additional covariates explored. Sections~\ref{sec:c6_marginal_models} and \ref{sec:c6_mvd} describe the marginal and dependence modelling of the process, respectively. In Section~\ref{sec:c6_results}, we use the model to explore how the properties of spatial extreme events have changed over time. Section~\ref{sec:c6_bts} outlines our novel spatio-temporal bootstrapping procedure, which is developed to account for strong intra-year climate oscillatory behaviour. Conclusions and a broader discussion are given in Section~\ref{sec:c6_conclusion}. All our code and instructions on how to access the data are available on GitHub\footnote{\url{https://github.com/dairer/Extreme-Irish-Winter-Temperatures}.}.

\section{Data}\label{sec:c6_data}

\subsection{Station data}
\label{sec:StationData}
\begin{figure}
\centering
  \includegraphics[width=.4\linewidth]{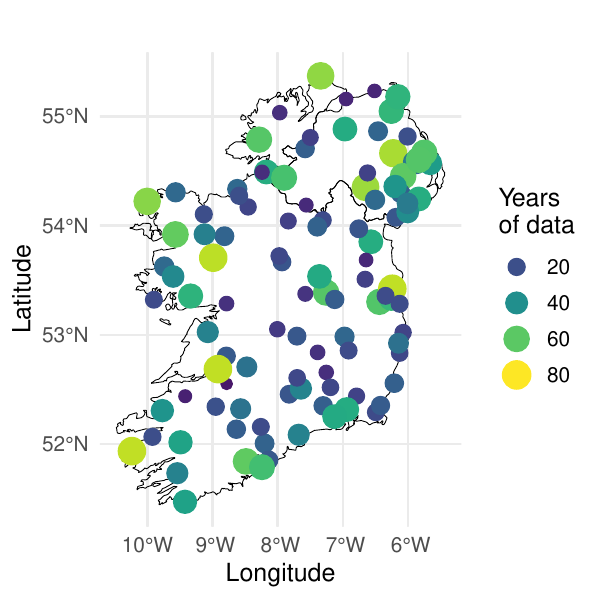}
  \caption{Station data locations with the amount of data indicated by colour and size.}
  \label{fig:obs_data_locs}
\end{figure}

We analyse observed temperature data from official Irish temperature recording stations rather than pre-processed data products, which may influence or smooth out extremal behaviour. Such smoothing can have a large impact on the magnitude of unusually large or small events \citep{Donat2014}. The data are compiled from two sources, with observational data for the Republic of Ireland provided by Met Éireann's data archive\footnote{Copyright Met Éireann. \url{https://www.met.ie/climate/available-data/historical-data}.}, while Northern Ireland sites were obtained through the Met Office Integrated Data Archive System \citep[MIDAS;][]{ukmo-midas}. The locations and lengths of data series of the 125 stations are shown in Figure \ref{fig:obs_data_locs}, where we only use stations with a least 5 years of data. We restrict our analysis to data from the period 1950--2022, although some stations have data prior to 1950, as these are the dates of our climatic covariates (discussed in Section~\ref{sec:c6_oscillation_covariates}).

We use daily minimum temperature data from the winter months (DJF). The DJF period accounts for more than 80\% of all the days with temperatures colder than their 1\% site-wise marginal quantile, a substantial increase relative to when using the equivalent 5\% quantile, suggesting that DJF accounts for the vast majority of extreme cold temperature days in Ireland. The data exhibit a substantial amount of missing data, with data availability greatly decreasing as we go back in time. See the Supplementary Material Section~1 for a plot showing the proportion of observed data by year for each station. The dataset contains some outliers that were not physically realistic. We removed the 93 temperatures that were greater than four sample standard deviations away from the mean minimum temperature on a given day. The total number of daily minimum temperatures was approximately $330,000$. 

{Before we proceed to use these data for statistical analysis, we must first address the cause and nature of the missing data. We intend to use all of the observed data, not simply the data simultaneously observed at all stations. Thus, we need to consider if the data are missing-not-at-random \citep{little2019statistical} as this can lead to bias, as was nicely illustrated for the type of models we consider by \cite{Richards2025}. It is important to clarify that much of what appears as missing data over the study period is not missing in any interesting statistical sense; rather, it's simply due to a station not yet being installed. Figure~1 in the Supplementary Material shows this is the primary cause of the \textit{missing data}, with massive variations in the start times. Furthermore, we believe missing values within a station's operational period are entirely due to standard issues concerning the movement of stations and the servicing of weather stations.  As temperature is relatively easy to monitor, it is exceedingly unlikely that the measuring equipment itself causes data to be missing. As the coastal stations have the longest records, due to being established for different purposes than inland stations, we have data missing-at-random, but not missing-completely-at-random \citep{little2019statistical}, which \cite{Richards2025} illustrates can lead to more variable inferences than if they met the more idealised conditions. We have decided to live with this limited inefficiency as the volume of missing data is so substantial that any type of infill method will either generate bias or huge computational complexity.}

\subsection{Spatial covariates}\label{sec:c6_covariates}
\subsubsection{Climate model output}\label{sec:clim_mod}
We explore the efficacy of exploiting physical information from climate model outputs as a spatial covariate. Climate models are mathematical representations of the physical processes driving weather and climate and represent our best understanding of these natural phenomena \citep{Giorgi2019}. They are computationally expensive, and so modelling the climate on a fine scale requires limiting ourselves to specific regions. Climate models are broadly run on two scales, general circulation climate models (GCMs) and regional climate models (RCMs). We use the output from an RCM to incorporate more detailed descriptions of the physics and spatial characteristics into the modelling of temperature extremes. We take data from the CLMcom (Climate Limited-area Modelling Community)-CLM-CCLM4-8-17 RCM combined with the ICHEC-EC-EARTH GCM \citep{CORDEX2019}. From the models, we have daily minimum temperature values over a 56-year period on a regular grid of 558 points over Ireland (corresponding to a $0.11^2$ degree resolution). For a given ensemble, we are required to select a so-called ``experimental configuration". Historical experiments are climate simulations for a period in which observational data exists, but they fundamentally differ from reanalysis data as they are not designed to match day-to-day weather observations, focusing instead on capturing general weather processes. For our chosen ensemble, this period covers 1951--2005. To simulate data beyond this period, it is necessary to make some assumptions about future climate change. However, our analysis relies exclusively on observed data to inform temporal trends in the extremal process, and so, does not require climate model output for informing temporal non-stationarity. We rely solely on the historical experiment for our spatial analysis. We do not use the climate model output as a direct covariate; rather, we first fit an extreme value model to the climate model output and use the parameter estimates as covariates in our final model of the observed data; details are given in Section~\ref{sec:c6_marginal_models}.

\subsubsection{Coastal distance}
There is an evident difference in temperature levels between coastal and inland areas in Ireland, especially so in winter \citep{Mateus2022}. This spatial effect is due to the strong influence of the Irish Sea and the Atlantic Ocean on Irish air temperatures. We are thus motivated to employ coastal proximity as a covariate to examine its efficacy in explaining the spatial distribution of cold extremes. We denote coastal proximity at site $\boldsymbol{s}$ as $C(\boldsymbol{s})$, calculated as the Euclidean distance from the nearest coastal point, using the \texttt{Simple Features} package in \texttt{R} \citep{sfpackage}.

\subsection{Temporal covariates}
\label{sec:temp-covariates}
\subsubsection{Overview}

In our analysis, we aim to disentangle the general, long-term trend of extremely cold winter temperatures from the large variations caused by highly variable climatic processes. To this end, we explore numerous climatic variables to help explain extreme, sudden cold snaps, which can be misleadingly interpreted as contradicting warming winters, i.e., covariates of the nature \cite{Koh2025} suggested in the discussion of \cite{Healy2023}. 

\subsubsection{Climatic oscillation covariates}\label{sec:c6_oscillation_covariates}

It is well understood that the jet stream 
plays a crucial role in shaping weather patterns and controlling the movement of storms and weather systems \citep{Hoskins2014}. It can lead to substantial shifts in weather patterns, including the occurrence of extreme events such as cold spells \citep{Stendel2021}. The jet stream is a high-altitude, fast-flowing air current that circumvents the northern hemisphere in the upper troposphere and lower stratosphere, which meanders in a wave-like pattern as it circumnavigates the globe. These waves are characterised by northward or southward oscillations. When these waves become amplified, we see a greater north-south displacement of air masses, resulting in the transport of air masses from different latitudes. When a deep southward dip forms in the jet stream, it can allow cold air to spill southward from the polar regions into lower latitudes, causing extremely cold temperatures \citep{ Thompson2021, Francis2012}. There is no simple measured covariate which captures the effects of the jet stream well, so climate scientists use measurable covariates which act as a proxy for the main features of the jet stream, with the most widely used being the North Atlantic Oscillation (NAO) and Arctic Oscillation (AO) \citep{Hall2015}. Both the NAO and AO are known to reflect the behaviour and position of the jet stream \citep{Gerber2009}. They have the greatest variability in winter months \citep{SenGupta2012}, which is when they have a substantial influence on weather patterns in the northern hemisphere \citep{Thompson1998}.  Although other atmospheric and oceanic patterns, such as high-pressure systems and sea surface temperatures, can also interact with the jet stream and influence the occurrence of cold weather events \citep{Screen2014}, these are generally considered more complex to model and measure through simple covariates. 

\begin{figure}[h]
    \centering
    \includegraphics[width = 12cm]{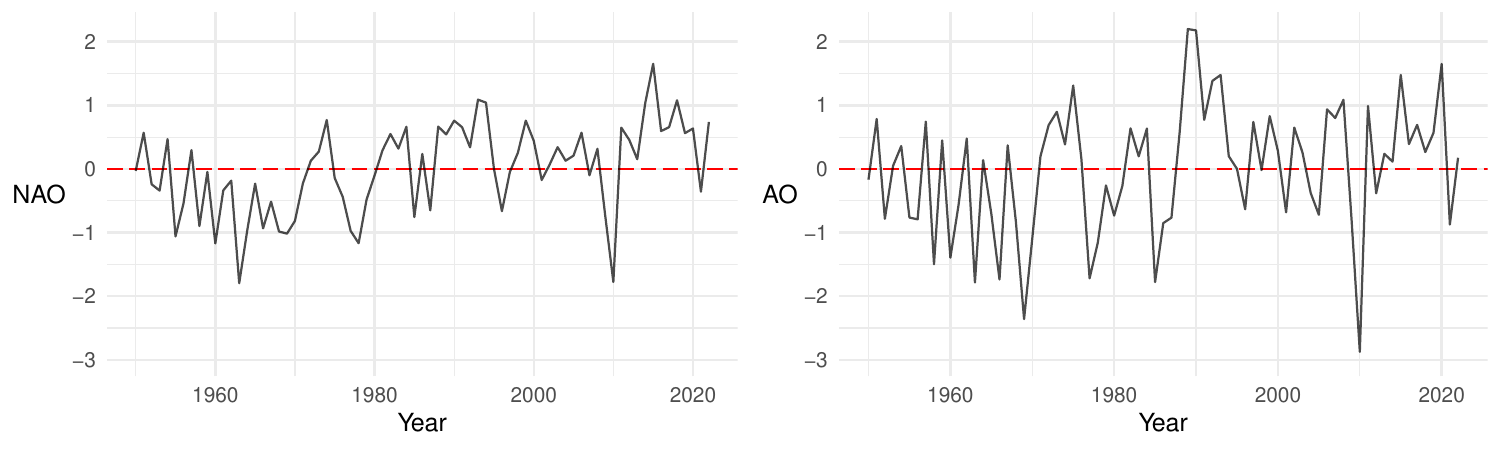}
    \caption[Yearly North Atlantic and Arctic Oscillation indices over the period 1950--2022.]{Winter (DJF) average indices for 1950--2022 of  (left) North Atlantic Oscillation (NAO)  and (right) Arctic Oscillation (AO).}
    \label{fig:ao_nao}
\end{figure}

We retrieve the NAO and AO covariates from the National Weather Service, Climate Prediction Center\footnote{Available at: \url{https://www.cpc.ncep.noaa.gov}.}, deriving monthly and winter (DJF) mean NAO and AO values (from the daily values) for 1950 onwards, with the latter shown in Figure~\ref{fig:ao_nao}. From this figure, we can see correlated behaviour between indices, and an upward trend in value for each index and large negative values for the 2009/10 winter, which appear more pronounced in the AO series. The winter of 2009/10 is well-known in Ireland for being extremely cold for a long period. The strong inter-annual winter variations in both covariates 
suggest that the key difference between Ireland's cold winter temperatures could possibly be explained better using these covariates than the longer-term smooth trend, which we introduce in Section~\ref{sec:c6_short_term_covariates}.

To help understand the features seen in Figure~\ref{fig:ao_nao}
some background detail on the two covariates is helpful. The NAO describes the state of the atmospheric pressure difference between the Icelandic Low and the Azores High in the North Atlantic. Similarly, the AO characterises changes in atmospheric pressure over the Arctic region \citep{Thompson1998}, i.e., the difference between the Arctic and mid-latitudes. The NAO and AO positive phases reflect that the jet stream is confined to higher altitudes, with a steady eastward flow.  Consequently, weather systems and air masses tend to move more smoothly and quickly across the mid-latitudes \citep{Deser2000}. Conversely, NAO and AO negative phases reflect a more meandering and meridional path of the jet stream. This allows the jet stream to dip southwards, pouring cold air masses onto northern Europe, and it also leads to a slower progression of weather systems, and subsequently more persistent weather regimes, such as a long-lasting cold spell \citep{sousa2018}.  Persistent and strong negative phases of the NAO and the AO have been linked with extremely low minimum air temperatures in  Ireland \citep{Mateus2022}. These properties of NAO and AO motivate us to use as a covariate for extreme cold winter temperatures in Ireland, although we do not use them together given their strong positive correlation. 

\subsubsection{Long- and short-term observational data product covariates}
\label{sec:c6_short_term_covariates}

\begin{figure}[!h]
    \centering\includegraphics[width = 10cm]{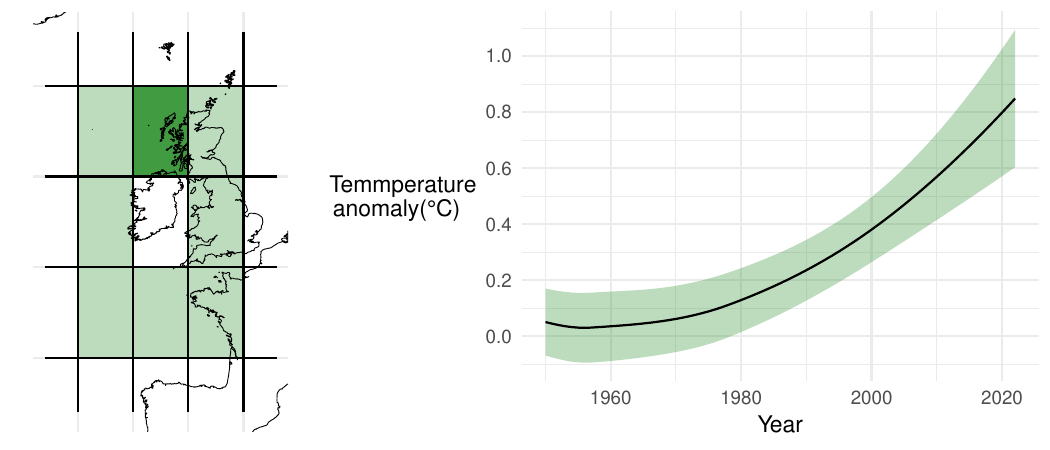}
    \caption{(left) Illustration of gridding of the HadCRUT5 dataset over northwest Europe centred on Ireland, with grid boxes surrounding Ireland lightly shaded and the grid box north of Ireland heavily shaded (left). (right) LOESS smoothed Irish winter temperature anomalies from HadCRUT5, with shaded regions indicating the pointwise 95\% confidence intervals.}
    \label{fig:c6_mean_temp_change}
\end{figure}

{As a temporal covariate, we use the HadCRUT5 observation-based data product \citep{Morice2021} to provide the monthly mean of daily average temperature anomalies across grid boxes over a coarse spatial scale, with Ireland represented by a single grid box. HadCRUT5 values over this monthly or winter scale represent large-scale spatial and temporal mean temperature variations and hence provide a climate change signal which we can use to help explain shifts in local-scale (daily) temporal behaviours of extreme temperatures. Figure~\ref{fig:c6_mean_temp_change} (left panel) shows the grid boxes of interest to us, with our primary focus being on the box centred on Ireland, but we will also consider the more northern grid box and the average over the adjacent grid boxes. We consider data from the grid point north of Ireland as a covariate separately in an attempt to isolate and investigate cold extremes driven by northerly airflows. Considering grid points adjacent to Ireland allows us to isolate the effect of large-scale atmospheric patterns that drive extreme winter minima. The monthly temperature values that HadCRUT5 provides arise from a combination of a land temperature anomaly dataset} \citep[CRUTEM5;][]{Osborn2021} {merged with a sea-surface temperature anomaly dataset} \citep[HadSST4;][]{Kennedy2019}. {An appealing feature of this data summary is that it accounts for and encapsulates the effects of all climate processes that affect temperature, e.g., including the jet stream, NAO and AO.} 

{We use only the winter months DJF of the HadCRUT5 data. We decompose the HadCRUT5 mean temperature anomaly data, $H^{I}_{m}(t)$, for month $m$ in year $t$ within the grid box containing Ireland (I), as }
\[
H^{I}_{m}(t)=H^{I}_{L,m}(t)+H_{R,m}^{I}(t),
\]
{where $H^{I}_{L,m}(t)$ and $H_{R,m}^{I}(t)$ respectively represent the long-term trend and short-term variations of the mean temperature anomaly data. Here $H^{I}_{L,m}(t)$ is defined to be smooth within and across years and $H_{R,m}^{I}(t)$ is simply the residuals from fitting $H^{I}_{L,m}(t)$, and hence it captures structure within and between sequential years. We also present average winter values of these covariates, which we denote by dropping the $m$ subscript. For the northern grid box and the average over the adjacent boxes, we replace the superscript I by N and A, respectively. We estimate $H^{I}_{L,m}(t)$ using the $\{H^{I}_{m}(t)\}$ data for the grid box $I$, by using the R language implementation of LOESS smoothing, taking the default span of $0.75$ and degree of $2$. The covariate $H^{I}_{L,m}(t)$, plotted in Figure~\ref{fig:c6_mean_temp_change}, shows there to be an increase of approximately $0.9^\circ$C in mean winter temperature over the period 1950--2022. }

\begin{figure}[h]
    \centering
    \includegraphics[width = \textwidth]{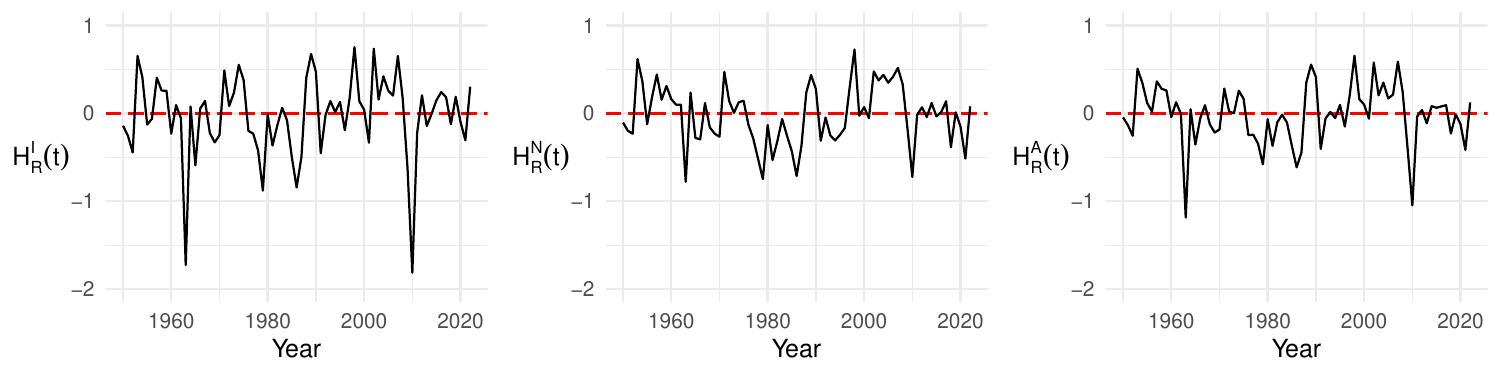}
    \caption{Time series of residual yearly average winter temperature anomalies:  (left) for the grid box over Ireland, (centre) for the grid box north of Ireland, and (right)  averaged over all grid boxes surrounding Ireland}.
    \label{fig:LOESS_smoothing_res_N_and_G}
\end{figure}

{The short-term variability of HadCRUT5 is given by the residuals from these LOESS smoothers, with Figure~\ref{fig:LOESS_smoothing_res_N_and_G} plotting the series of winter values of $\{H_{R}^{I}(t)\},  \{H_{R}^{N}(t)\}$ and $\{H_{R}^{A}(t)\}$. Positive residuals correspond to anomalously cold mean winter temperatures relative to the long-term increasing winter temperature, which provides a represent of large-scale climatic oscillatory patterns around Ireland.  In all three of these residual time series, there is a large positive deviation in the winter of 2009/10, a feature separately identified for the NAO and AO indices in Figure~\ref{fig:ao_nao}.}

\subsubsection{Presentation of short-term climatic variability (SCV)}\label{sec:scv}
{Since NAO, AO, $H^I_{R,m}(t)$, $H_{R}^{N}(t)$ and $H_{R}^{A}(t)$ are highly variable covariates, the results of our analysis will change considerably depending on their values. For the sake of interpretability, in subsequent analyses, we report results for three different $\tau$-th quantile levels of these covariates:  $\tau= 0.1, 0.5$ and $0.9$, and we add the associated $\tau$ values as a superscript to the covariate. We refer to these as low, median, and high phases of short-term climatic variability (SCV), respectively. This form of presentation provides a general interpretation of different levels of extremity over the different phases of SCV. The high phase of the SCV covariate $H^{I}_{R,m}(t)$ corresponds to a large negative phase of NAO and AO and subsequently a pronounced meandering of the jet stream which we know is linked to more extreme cold spells in Northern Europe, as discussed in Section~\ref{sec:c6_oscillation_covariates}.}

{We explore a set of covariates to assess how variations in large-scale atmospheric circulation patterns influence local temperature extremes and their SCV. This set varies over time $t$, and is defined as}
\[
\mathcal{Z}(t) = \{ \varnothing , ~M^{I}_{R,m}(t), ~M^{I}_{R}(t),  ~\text{NAO}(t),  ~\text{NAO}_R(t), ~\text{NAO}_{m}(t), ~\text{NAO}_{R,m}(t), ~\text{AO}(t), ~\text{AO}_R(t), ~\text{AO}_{m}(t), ~\text{AO}_{R,m}(t), ~M_R^{A}(t), ~M_R^{N}(t)\}.
\]
{The covariate $M_R^{A}(t)$ refers to the average HadCRUT5 values over the grid points adjacent to Ireland, excluding the grid point over Ireland, and $M_R^{N}(t)$ refers to the average HadCRUT5 values over the grid points above Ireland. The covariates $\text{NAO}(t)$ and $\text{AO}(t)$ are the NAO and AO indices at time $t$. In each case, the subscript $R$ indicates that the long-term trend linked to the smooth temperature anomalies, e.g., $H_L^I(t)$ as seen in Figure~\ref{fig:c6_mean_temp_change}, has been removed. This ensures that the covariates are orthogonal to the long-term mean temperature trend and represent SCV. A subscript $m$ indicates monthly values, whereas no subscript $m$ indicates that the average value over the winter months was taken for each year. We also investigated monthly values of $M_R^{N}(t)$ and $M_R^{A}(t)$. However, as compared to $M^{I}_{R}(t)$, their performance was generally worse in each case, so we do not include them for brevity. To simplify subsequent notation, we denote ${\mathcal{Z}(t): =\{z_j(t):\:j=0,1,\dots, 12\}}$, where $z_0(t)=\varnothing$, $z_1(t)=M^{I}_{R,m}(t)$, and so on. }

\section{Marginal models}\label{sec:c6_marginal_models}
\subsection{Overview and strategy}
\label{sec:marginal_strategy}
Let $X_{\text{o}}(t,\boldsymbol{s})$ denote the negated observed minimum daily temperature at time $t$ and site $\boldsymbol{s}$ during winter months, and let $X_{\text{c}}(t,\boldsymbol{s})$ be the negated minimum daily temperature from the climate model at time $t$ and site $\boldsymbol{s}$. The subscripts $o$ and $c$ are used throughout, referring to observation and climate model processes, respectively.
Here $t \in \mathbb{N}$ indexes winter days within and across years and $\boldsymbol{s} \in \mathcal{S} \subset \mathbb{R}^2$, where $\mathcal{S}$ denotes Ireland, with $\boldsymbol{s}$ corresponding to the vector of latitude and longitude. We have data on the two processes at $\mathcal{S}_o\subset \mathcal{S}$ and $\mathcal{S}_c\subset \mathcal{S}$ and at times
$\mathcal{T}_o$ and 
$\mathcal{T}_c$ respectively. Critically, for our analysis, since we are modelling negated temperatures, higher quantiles correspond to colder temperatures. 

{With several different sources of information on the extremal temperature process, there is a range of possible ways to integrate this information when modelling the process. Methodology for the data fusion of spatio-temporal analysis of data from multiple such sources has been developed widely in the context of statistical downscaling, with examples in the extreme values context such as \cite{Berrocal2010} and \cite{cuba2025preprint}. Much of the focus of those methods is exclusively on modelling the marginal distribution of the process at a given site, rather than the full spatial dependence structure.  Instead, we present our related approach, which is largely motivated by the successful analysis of extremal data in \cite{Healy2023}, though we note that a number of the discussion contributions of that paper, namely by \cite{Cuba2025} and \cite{castillo-mateo2025}, identify strong parallels between our methodology and statistical downscaling. A further reason we do not follow a stronger data fusion approach for this particular data analysis is that we find that our other proxy data source, e.g., climate model outputs, does not provide a good description for extremely cold winter temperature behaviour; see Section~\ref{sec:c6_marg_tail}.}

We model the marginal distribution function
$F_o$, of $X_o(t,\boldsymbol{s})$, to vary smoothly 
over $(t,\boldsymbol{s})\in \mathcal{T}\times \mathcal{S}$
using a threshold-based model, determined by a spatially varying threshold,
denoted by $u_o(\boldsymbol{s})$. This threshold is selected such that the probability of exceeding it for any $(t,\boldsymbol{s})$ is small, so values that exceed the threshold are deemed to be extreme. We separately model the behaviour of the non-extremal temperatures below a threshold, denoted as the \textit{bulk} model, and the extremal temperatures above, denoted as the \textit{tail} model. The combined model of the entire distribution is then constructed to ensure
a continuous transition from the bulk to the tail model for all $(t,\boldsymbol{s})$. Specifically, we take
\begin{equation}
F_o(y;t,\boldsymbol{s})= 
\left\{
\begin{array}{ll}
    F_B(y;t,\boldsymbol{s}) & \mbox{ if } y<u_o(\boldsymbol{s})\\
    F_B(u_o(\boldsymbol{s});t,\boldsymbol{s})+ [1-F_B(u_o(\boldsymbol{s});t,\boldsymbol{s})]F_T(y-u_o(\boldsymbol{s});t,\boldsymbol{s})
    & \mbox{ if } y\ge u_o(\boldsymbol{s}),    
    \end{array}
\right.
\label{eqn:joinedmarginal}
\end{equation}
where $F_B$ is a spatio-temporal distributional model for the full distribution, but which we only use for the bulk, i.e., up to a spatial threshold $u_o(\boldsymbol{s})$ only, and the conditional distribution of spatio-temporal excesses of $X_o(t,\boldsymbol{s})$ over $u_o(\boldsymbol{s})$ is given by $F_T$.

As there is no theoretical basis 
for choosing any one parametric form for $F_B$ and $u_o(\boldsymbol{s})$, we use a semi-parametric model approach, which uses quantile regression, see Sections~\ref{sec:c6_bulk}
and \ref{sec:c6_threshold_selection}
respectively. In contrast, based on standard extreme value asymptotic arguments, we take $F_T$ to be the generalised Pareto distribution (GPD) \citep{Davison1990}, which has distribution function
\begin{equation}
 F_T(y; \sigma,\xi)=
	1 - (1 + \xi y / \sigma )_{+}^{-1/\xi} ,
    \label{eqn:GPD}
\end{equation}
for $y>0$, with a shape parameter $\xi \in \mathbb{R}$ and a scale parameter $\sigma>0$, with the notation $x_+=\max(x,0)$.  This distribution is denoted by a GPD$(\sigma,\xi)$, and has a heavy and unbounded upper tail when $\xi>0$, it is the exponential distribution when $\xi=0$, and it has a finite upper bound to the distribution when $\xi<0$. The form of the distribution function when $\xi=0$ is obtained by taking the limit as $\xi\rightarrow 0$. In Section~\ref{sec:c6_marg_tail}, we present the covariate formulations of the GPD that we explored.

As is implicit in formulation~\eqref{eqn:joinedmarginal}, an important aspect of our modelling of the threshold is that we take it to be constant over time, a choice that we justify in Section~\ref{sec:c6_threshold_selection}. To inform the modelling stages for $F_B$ and $F_T$, we use the spatial covariates from 
Section~\ref{sec:c6_covariates}
and the temporal covariates from Section~\ref{sec:temp-covariates}, 
which capture both long- and short-term modes of climatic variability acting over the study domain. Uncertainty quantification of the modelling process for $F_B$ and $F_T$ is handled via a bespoke bootstrap method, presented in Section~\ref{sec:c6_bts}. That method is designed to deal with the combination of highly temporally dependent extremal data and the substantial missing nature of the data features of the spatio-temporal station data.

\subsection{Marginal data analysis}
\label{sec:MargData}
\subsubsection{A model for the bulk of the distribution}\label{sec:c6_bulk}
{We describe a model for the bulk component of expression~\eqref{eqn:joinedmarginal}, i.e., $F_{B}(y;t,\boldsymbol{s})$. We do this by using a semi-parametric estimate of the entire distribution function of $X_o(t,\boldsymbol{s})$, but which is only used for $F_o$ below
$u_o(\boldsymbol{s})$. We make this inference on $F_B$ implicitly by estimating a range of spatially and temporally varying $\tau$-th quantiles of $X_o(t,\boldsymbol{s})$, which we denote as $q_o^{(\tau)}(t,\boldsymbol{s})$, for a grid of $\tau\in  \{0.01, 0.02, \dots, 0.99\}$, for all $t \in \mathcal{T}$ and $\boldsymbol{s} \in \mathcal{S}$. We achieve this quantile regression through using the asymmetric Laplacian distribution (ALD), a distribution often applied in quantile regression \citep{Youngman2019}, parametrised by a location parameter (the $\tau$-quantile, $q_o^{(\tau)}(t,\boldsymbol{s})$), along with a scale, and an asymmetry parameter. Following this step, to complete our model inference, we use a cubic interpolation spline to give a continuous estimate over $\tau$, for each $t$ and $\boldsymbol{s}$. With this set-up, we have that our estimate of $F_B$ is obtained through the property that $F_{B}\left(q_o^{(\tau)}(t,\boldsymbol{s});t,\boldsymbol{s}\right)=\tau$ for all $\tau\in (0,1)$. }

We explored several potential linear model covariate formulations for the location parameter function, $q_o^{(\tau)}(t,\boldsymbol{s})$, of the ALD in our quantile estimation. We present these models in Table~\ref{tab:cold_extremes_quant_reg_models}, along with their corresponding RMSE values for a pair of cross-validation metrics, namely spatio-temporal cross-validation (ST-CV) and $n$-fold cross-validation ($n$-CV), which are detailed in the Supplementary Material, Section~2. Models~\textit{a.}-\textit{d.} follow the structure of the models considered by \cite{Healy2023}. Specifically, model~\textit{a.}, can be considered as the base model where a constant quantile is estimated over space and time for each $\tau$;  model~\textit{b.} includes the covariate for the smoothed temperature anomalies over Ireland, $M^I_L(t)$, and hence allows for temporal non-stationarity; model~\textit{c.}, has the covariate being the empirical sample $\tau$-th quantiles of the climate model output, $X_c(\boldsymbol{s})$ at the closest climate model grid point to each observed station, allowing the quantile estimates to vary spatially; and model~\textit{d.} includes both the covariates of models~\textit{b.}, and \textit{c.}, allowing for both spatial and temporal non-stationarity. Model~\textit{e.} is equivalent to model~\textit{d.} with the addition of the coastal proximity covariate $C(\boldsymbol{s})$, which was not required for the bulk model of hot summer temperatures. The RMSE values in Table~\ref{tab:cold_extremes_quant_reg_models} show that of the models without SCV covariates, i.e., models~\textit{a.}-\textit{e.}, show a progressive steady improvement of fit over all these five models, with a performance markedly improved when incorporating the coastal proximity covariate $C(\boldsymbol{s})$. Our novel inclusion of SCV covariates, not considered by \cite{Healy2023}, in models \textit{f.}-\textit{h.} makes further reductions in the RMSE cross-validation metrics relative to model~\textit{e.}. Namely, we explore the inclusion of $H_{R,m}^I(t)$, $\text{AO}_{R,m}(t)$ and $\text{NAO}_{R,m}(t)$ in models \textit{f.}-\textit{h.} respectively, with each of these covariates being constant for all $t$ in the respective month $m$. In particular, a substantial improvement in RMSE is found with the inclusion of $H_{R,m}^I(t)$, the residuals from the LOESS smoothing of temperature anomalies during winter months over Ireland. Hence, we focus on model~\textit{f.} for the bulk of the distribution.

\begin{table}
\centering 
\begin{tabular}{l|lcc} 
& \textbf{Model structure for $\hat{q}_{o}^{(\tau)}(t, \boldsymbol{s})$} & {\textbf{ST-CV}} & {\textbf{$n$-CV}}\\*
    \hline
    \textit{a.} & $\beta^{(\tau)}_0$ & 2.019 & 2.028 \\*
    \textit{b.} & $\beta^{(\tau)}_0 + \beta^{(\tau)}_1 H^{\text{I}}_L(t)$ & 1.991 & 2.028 \\ 
    \textit{c.} & $\beta^{(\tau)}_0 + \beta^{(\tau)}_1 q^{(\tau)}_{c}(\boldsymbol{s})$ & 1.898 & 1.912 \\*
    \textit{d.} & $\beta^{(\tau)}_0 + \beta^{(\tau)}_1 q^{(\tau)}_{c}(\boldsymbol{s}) + \beta^{(\tau)}_2 H^{\text{I}}_L(t)$ & 1.864 & 1.891 \\ 
    \textit{e.} & $\beta^{(\tau)}_0 + \beta^{(\tau)}_1 q^{(\tau)}_{c}(\boldsymbol{s}) + \beta^{(\tau)}_2 H^{\text{I}}_L(t) + \beta^{(\tau)}_3 C(\boldsymbol{s})$ & 1.764 & 1.756 \\*
    \rowcolor{green!15}\textit{f.} & $\beta^{(\tau)}_0 + \beta^{(\tau)}_1 q^{(\tau)}_{c}(\boldsymbol{s}) + \beta^{(\tau)}_2 H^{\text{I}}_L(t) + \beta^{(\tau)}_3 C(\boldsymbol{s})  + \beta^{(\tau)}_4 H^{\text{I}}_{R,m}(t)$ & \textbf{1.218} & \textbf{1.221} \\*
    \textit{g.} & $\beta^{(\tau)}_0 + \beta^{(\tau)}_1 q^{(\tau)}_{c}(\boldsymbol{s}) + \beta^{(\tau)}_2 H^{\text{I}}_L(t) + \beta^{(\tau)}_3 C(\boldsymbol{s})  + \beta^{(\tau)}_4 \text{AO}_{R,m}(t)$ & {1.451} & {1.446} \\
    \textit{h.} & $\beta^{(\tau)}_0 + \beta^{(\tau)}_1 q^{(\tau)}_{c}(\boldsymbol{s}) + \beta^{(\tau)}_2 H^{\text{I}}_L(t) + \beta^{(\tau)}_3 C(\boldsymbol{s})  + \beta^{(\tau)}_4 \text{NAO}_{R,m}(t)$ & {1.430} & {1.433} \\\hline
  
\end{tabular}
\caption{Cross-validation (RMSE) on the quantile regression analysis for the bulk of the distribution.}\label{tab:cold_extremes_quant_reg_models}
\end{table}

\begin{figure}[h]
    \centering
    \includegraphics[width = 14cm]{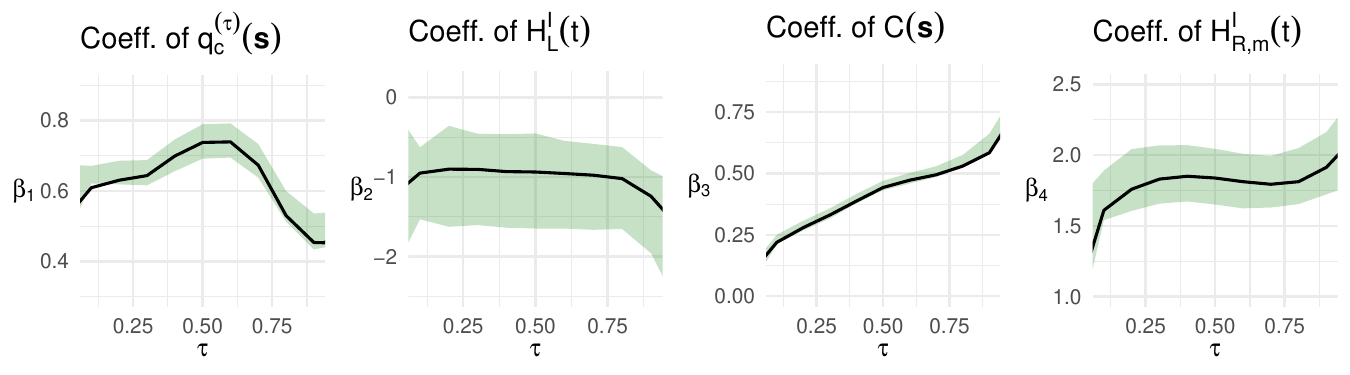}
    \caption[Estimates of bulk model~\textit{f.}~coefficients.]{Estimates of bulk model~\textit{f.}~coefficients $\beta^{(\tau)}_1$, $\beta^{(\tau)}_2$, $\beta^{(\tau)}_3$, and $\beta^{(\tau)}_4$ of $q_c^{(\tau)}(\boldsymbol{s)}$ (corresponding climate output quantile), $H^I_L(t)$ (smoothed temperature anomalies over Ireland), $C(\boldsymbol{s)}$ (coastal distance), and  $H^I_{R,m}(t)$ (short-term behaviour of the temperature anomaly data, i.e., SCV) over a range of quantiles $\tau$. In each case, the shaded region indicates bootstrap-based pointwise $95\%$ confidence intervals. Higher quantiles correspond to colder temperatures.}
    \label{fig:blk_qnt_reg_params}
\end{figure}

The estimated $\beta^{(\tau)}$-coefficients of model \textit{f.} are shown in Figure \ref{fig:blk_qnt_reg_params}, along with 95\% uncertainty intervals based on 200 spatio-temporal bootstrap samples described in Section~\ref{sec:c6_bts}. The estimates of $\beta_1^{(\tau)}$ show that the climate model does not provide a perfect description of the station data, as the estimates differ significantly from $1$ and change with $\tau$. This suggests that the chosen climate model output is not sufficient in and of itself in representing daily winter minima. However, the climate model is still informative here, especially for central quantiles (i.e., $0.4 < \tau < 0.6$), where it is closest to 1. The climate model covariate appears to become less informative towards the extremes of the process as we see the effect of the $q_c^{\tau}$ falling off at both tails of the distribution, particularly for $\tau > 0.8$ corresponding to the coldest 20\% of daily minimum temperatures. For the majority of the bulk of the distribution, we see that $\beta_2^{(\tau)}\approx-1$. Since we are modelling negated winter minima, $X_o(t,\boldsymbol{s})$, a coefficient value of $-1$ indicates that extremely cold winter temperatures are warming at a similar rate to the mean winter temperatures, so for the majority of quantiles in the bulk of the distribution this property holds. However, the effect of $H^I_L(t)$ is strengthening at higher quantiles ($\tau > 0.75$), i.e., $\beta_2^{(\tau)}$ decreased below $-1$, suggesting that extremely cold daily winter temperature events are warming at a faster rate than mean daily winter temperatures over Ireland. The effect of coastal proximity increases almost linearly with quantiles $\tau$, emphasising that it becomes increasingly important relative to the climate model covariate $q_c^{(\tau)}(\boldsymbol{s})$, as we move to colder events. Finally, as estimates of $\beta_4^{(\tau)}$ are statistically significantly different from 0 over $\tau$, this shows that large-scale oscillations are useful for explaining the daily minimum temperatures, even after the long-term trend and spatial features have been accounted for. As $\beta_4^{(\tau)}$ is largest for $\tau > 0.8$, this covariate's importance is strongest for extremely cold days.

\subsubsection{Threshold selection and exceedance probabilities}\label{sec:c6_threshold_selection}

Before setting out our model and inference process for the threshold, first we explain our choice of the threshold taken to be constant over time, i.e., as $u_o(\boldsymbol{s})$ in formulation~\eqref{eqn:joinedmarginal}. There are a range of reasons we adopt this strategy, theoretical and empirical, with much of these discussed in detail in our companion paper on extremely hot summer temperatures \citep{Healy2023}. They argued that if long-term trends (the focus of the analysis) are small relative to natural variability in the data, it is inappropriate to let a subjectively chosen threshold vary over time or with temporal covariates, because doing so could bias the inference of key parameters of the extreme value model. 

In our context, this implies that we have to consider both (i) long-term trends  (identified in Sections~\ref{sec:c6_oscillation_covariates}), and (ii) substantial year-to-year variability in cold winter temperatures (identified in Section~\ref{sec:c6_short_term_covariates}), relative to the daily winter variability. To investigate the evidence for features (i) and (ii), we adopt a 96\% marginal threshold at each site and identify extreme temperature days as the exceedances of this level. We examine the temporal occurrence rates of these exceedances across all sites and 73 years of data. The winter of 2009/10 has a disproportionately large percentage of extreme events, with $\sim17$\% of exceedances occurring in that winter. However, other than that, winter exceedances are relatively evenly spread across space and time. These findings were repeated over a range of $\tau$-th quantiles, for $\tau>0.9$. To alleviate the impact of the winter of 2009/10 on the estimation of the threshold, we could use a different threshold estimate during 2009/10 from other years. However, a step function in the threshold for that winter greatly reduces the interpretability and parsimony of our model. Instead, we consider it to be preferable to keep the threshold constant over time and capture the rate of threshold exceedance and the parameters of the excess distribution through the physically interpretable covariates described in Sections~\ref{sec:c6_oscillation_covariates} and \ref{sec:c6_short_term_covariates}.

\begin{figure}[h]
    \centering
    \includegraphics[width = 11cm]{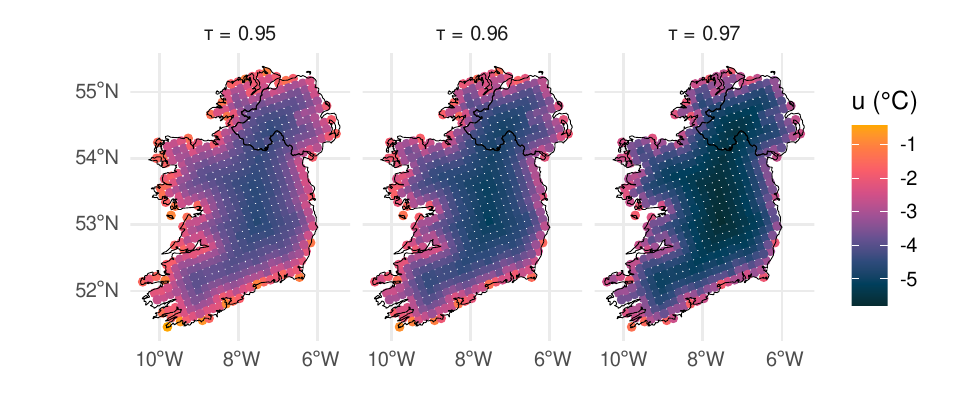}
   \caption[Negated estimates of threshold $u_o^{(\tau)}(\boldsymbol{s})$.]{Estimates of the negated threshold $u_o^{(\tau)}(\boldsymbol{s})$ for $\tau\in \mathcal{S}$, with values in $^{\circ}C$: (left to right) $\tau =0.95,0.96,0.97$. }
    \label{fig:threshold_plots}
\end{figure}

To estimate the threshold $u_o(\boldsymbol{s})$, we use spatial $\tau$-th quantile regression using the ALD approach as detailed in Section~\ref{sec:c6_bulk}. For the $\tau$-th quantile at site $\boldsymbol{s}$ for $X_o(t,\boldsymbol{s})$ we use the threshold formulation:
\begin{align}\label{eq:thresh_mod}
    u^{(\tau)}_o(\boldsymbol{s}) = \beta_0^{(\tau) } + \beta_1^{(\tau) }q^{(\tau)}_c(\boldsymbol{s}) + \beta_2^{(\tau) } C(\boldsymbol{s}),
\end{align}
which is equivalent to our chosen model \textit{f.}~(in Table~\ref{tab:cold_extremes_quant_reg_models}), simply without temporal covariates. Our primary analysis is performed using the $\tau=0.96$ quantile, motivated by exploratory analysis using a shape parameter stability plot \cite{Coles2001}; however, we also calculate results for $\tau=0.95, 0.97$ quantiles to assess the sensitivity to threshold selection. Estimates of $u^{(\tau)}_o(\boldsymbol{s})$ using regression~\eqref{eq:thresh_mod} for each $\tau \in \{0.95,0.96,0.97\}$ are given in  Figure~\ref{fig:threshold_plots}. This figure shows that inland areas of Ireland experience colder winter temperatures by 3-5$^{\circ}C$ than coastal regions, with the coastal transition region decreasing in area with colder temperature events, i.e., as $\tau$ increases, as identified in Figure~\ref{fig:blk_qnt_reg_params} (panel 3). 

\begin{figure}[h]
    \centering
    \includegraphics[width = 14cm]{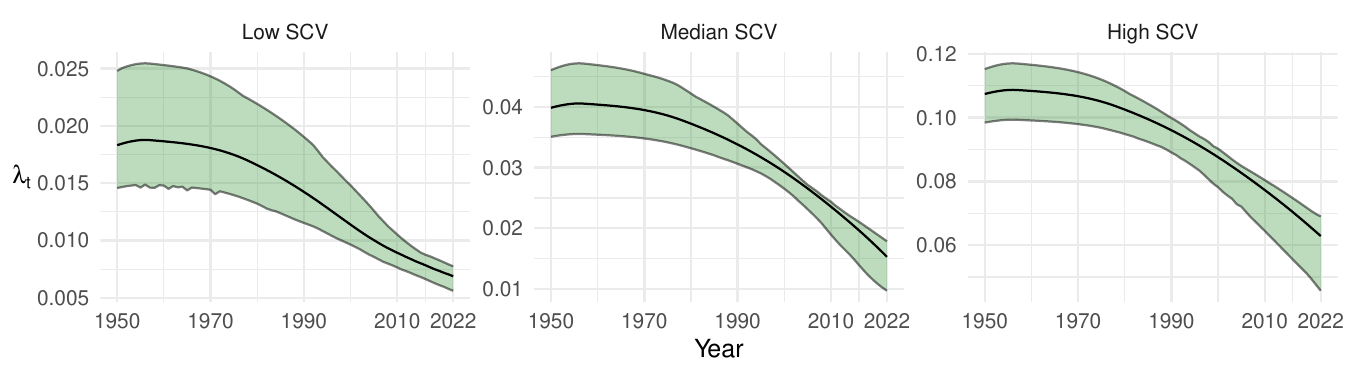}
    \caption[Threshold exceedance probability $\lambda_o(t)$ from 1950 to 2022.]{Spatially averaged exceedance probabilities $\lambda^{(\tau)}_o(t)$ of the threshold for $t$ from 1950 to 2022 averaged over $\mathcal{S}$ for $\tau=0.96$:    
    for low (left), median (centre) and high (right) phases of SCV. The shaded regions indicate bootstrap-based pointwise $96\%$ confidence intervals for $\lambda^{(\tau)}_o(t)$.}
    \label{fig:cold_ex_blk_qnt_thresh_ex}
\end{figure}

We can now use our estimate of the bulk of the distribution $F_B(y;t,\boldsymbol{s})$ from Section~\ref{sec:c6_bulk} with our estimate $u^{(\tau)}_o(\boldsymbol{s})$ of the threshold $u_o(\boldsymbol{s})$, based on a particular choice of $\tau$, to give an estimated threshold exceedance probability
\[
\lambda^{(\tau)}_o(t,\boldsymbol{s}):=
1-F_B(u_o^{(\tau)}(\boldsymbol{s});t,\boldsymbol{s}).
\]
To illustrate the temporal changes in the threshold exceedance rate, we plot the spatial average exceedance rate over Ireland, i.e., 
\[
\lambda^{(\tau)}_o(t)=
\int_{\boldsymbol{s}\in \mathcal{S}}
\lambda_o^{(\tau)}(t,\boldsymbol{s})\,d\boldsymbol{s}/|\mathcal{S}| \mbox{ for }t\in \mathcal{T}_o,
\]
in Figure~\ref{fig:cold_ex_blk_qnt_thresh_ex}, where we show $\lambda^{(\tau)}_o(t)$ against time for the low, median, and high phases of SCV separately, for $\tau=0.96$. Thus, if the process $X_o(t,\boldsymbol{s})$ was stationary over time, $\lambda^{(\tau)}_o(t)=0.04$. The estimates show a decreasing rate of daily minimum temperatures colder than the threshold over the period 1950--2022, for each phase of SCV, with these rates changing by a factor of at most $4$ over years. Critically, though, the phase of SCV is much more influential on the rate, with changes by a factor of up to 10 in the rate from lowest to highest SCV illustrative values. In particular, the nominal average rate of $0.04$ is only achieved in 2022 in conjunction with a high level of SCV, and in 1950 with median phase of SCV. For these estimates, pointwise 96\% confidence intervals, based on our spatio-temporal bootstrap method presented in Section~\ref{sec:c6_bts}, show a shrinking in uncertainty in the estimates as time progresses. {This is likely due to the approximate binomial sampling distribution of the estimators. Consequently, they have variances which are proportional to $\lambda^{(\tau)}_o(t)$, which are themselves decreasing due to the warming of cold winter temperatures. Critically, by the nature of the construction of our model, this shrinkage does not reflect any changes in the number of active station recording temperatures.}

\subsubsection{Tail model}\label{sec:c6_marg_tail}

Conditionally on $X_o(t, \boldsymbol{s}) > u_o(\boldsymbol{s})$, we are in the tail region of interest, and model the threshold excesses $X_o(t, \boldsymbol{s}) - u_o(\boldsymbol{s})$. As outlined in Section~\ref{sec:marginal_strategy}, for each site $\boldsymbol{s} \in \mathcal{S}$, we model $F_T$, the distribution function of excesses of the threshold $u_o(\boldsymbol{s})$, 
as GPD$(\sigma,\xi)$ and consider covariates for each of the parameters. Given our findings for the bulk model for large $\tau$ in Section~\ref{sec:c6_bulk}, a key difference of our approach, relative to that of \cite{Healy2023}, is that here the central focus is on incorporating covariates that reflect the SCV effects.

It is typical in fitting the GPD to non-identically distributed environmental applications to take the shape parameter to be constant over space and time due to difficulty in estimating the parameter \citep{Chavez-Demoulin2005}. Here, we assess the evidence for this for both the observational and climate model data. We explore if there is evidence that their respective shape parameters $\xi_o(\boldsymbol{s})$ and $\xi_c(\boldsymbol{s})$ are constant over space, i.e., if they are equal to unknown values $\xi_o$ and $\xi_c$ for all $\boldsymbol{s} \in \mathcal{S}$ respectively. We do not explore temporal changes in these parameters due to the short duration of our sampling window and the subtle and small changes in levels over this period relative to other sources of variation. To perform such tests, we fit a GPD model in which the scale and shape parameters vary by site under the alternative hypothesis, whereas under the null, the respective shape parameters are constrained to be $\xi_o$ and $\xi_c$ across all sites. We ignore temporal dependences in constructing a likelihood and perform the likelihood ratio test for each site $\boldsymbol{s} \in \mathcal{S}_o$ and $\mathcal{S}_c$ respectively. Results are presented for the observed data. Specifically, we found that over 91\% of sites had a Wald test statistic below $0.05$, with 95\% of sites below $0.06$, i.e., well below the 3.84 critical value of a 95\% test.  Furthermore, there was no clear spatial pattern in those sites with a test statistic higher than $0.05$. Thus, we choose to keep a constant shape parameter. Using the fit with $\hat{\xi}_c$ gives us our final covariate, $\sigma_c(\boldsymbol{s})$ for $\boldsymbol{s} \in \mathcal{S}_c$, for modelling the scale parameter of $X_o(t, \boldsymbol{s})$. The covariates $\sigma_c(\boldsymbol{s})$ for the different thresholds we consider are given in the Supplementary Materials, Section~3.

We now focus on investigating numerous choices of models for $\sigma_o(t, \boldsymbol{s})$. We present these models in eight different groups of formulations, labelled $A-H$, each with variants labelled by a subscript, see Table~\ref{tab:gpd_model_groups}. Models in groups $A$ and $B_0-H_0$ are of the type considered by \cite{Healy2023}, whilst models $B_j-H_j$, for $j=1, \ldots, 13$ are novel and address the property that Irish winter minimum temperatures are highly variable and strongly associated with large-scale climatic oscillations, as discussed in Section~\ref{sec:c6_intro}, and so these models incorporate the SCV covariates $z_j(t)$ defined in Section~\ref{sec:scv}, where we consider the full set of covariates $\mathcal{Z}(t)$, unlike the restricted set in the bulk model. Recall that for each $j$, $z_j(t)$ is constant for all $t$ in a given month, or a winter, depending on whether the covariate is indexed by $m$ or not, respectively. Our group $A$ of models is a set of three temporally invariant ``base" models which we include for comparison against our proposed models; group $B$ assumes spatial stationarity but incorporates the smooth temporal mean trend; groups $C$ and $D$ borrow spatial information from the climate model output in addition to the features in group $B$, whereas $D$ also explores a coastal proximity effect; group $E$ expands group $D$ to additionally explore the interaction between the coast and long-term temporal trends; groups $F$ and $G$ do not borrow information from the climate model output, instead they rely on coastal proximity to describe all spatial non-stationarity; whilst group $H$ has an interaction in space and time but no main spatial term. This gives 94 models in total, all of which, along with their cross-validation scores, can be seen in the Supplementary Material, Section~4. 

\begin{table}[!htbp]
 \centering
  \begin{tabular}{c|l}
  \textbf{Group} & \textbf{Model structure for} $\log \sigma_{o}(t,
  \boldsymbol{s})$ \textbf{for} $j = 0,\dots,13$\\
    \hline
      $A_0$ & $\beta_0$\\ 
      $A_1$ & $\beta_0+ \beta_1\sigma_c(\boldsymbol{s})$\\ 
      $A_2$ & $\beta_0+ \beta_1C(\boldsymbol{s})$\\\hline
      $B_j$ & $\beta_0 + \beta_1M^{I}(t) + \beta_2 z_j(t)$\\ 
      $C_j$ & $\beta_0 + \beta_1\sigma_c(\boldsymbol{s}) + \beta_2M^{I}(t) + \beta_3 z_j(t)$\\ 
      $D_j$ & $\beta_0 + \beta_1\sigma_c(\boldsymbol{s}) +\beta_2C(\boldsymbol{s}) + \beta_3M^{I}(t) + \beta_4 z_j(t)$\\ 
      $E_j$ & $\beta_0 + \beta_1\sigma_c(\boldsymbol{s}) + \beta_2M^{I}(t) + \beta_3M^{I}(t)C(\boldsymbol{s}) + \beta_4 z_j(t)$\\ 
      $F_j$ & $\beta_0 + \beta_1C(\boldsymbol{s}) + \beta_2M^{I}(t) + \beta_3 z_j(t)$\\ 
      $G_j$ & $\beta_0 + \beta_1C(\boldsymbol{s}) + \beta_2M^{I}(t) + \beta_3M^{I}(t)C(\boldsymbol{s}) + \beta_3 z_j(t)$\\ 
      $H_j$ & $\beta_0 + \beta_1M^{I}(t) + \beta_2M^{I}(t)C(\boldsymbol{s}) +\beta_4 z_j(t)$ \\ \hline
    \end{tabular}
    \caption{Groups of covariate model log-linear formulations for the GPD scale parameter $\sigma_o(t,\boldsymbol{s})$. In each group, $z_j(t)$ (for $j>0$) represents all additional temporal covariates to capture SCV, with the associated covariates being defined in Section~\ref{sec:c6_oscillation_covariates}.}
    \label{tab:gpd_model_groups}
\end{table}

We use RMSE and CRPS jointly to identify the best-performing models and most effective covariates. Following this, we use CRPS to choose whether to take a given covariate on a monthly or yearly scale. For the cross-validation metrics presented in Table~\ref{c6:tab_cv_gpd} (and in Supplementary Material Table~1), empirical quantiles used for calculating RMSE were estimated on yearly blocks of data for each site (in order to have sufficient data for estimating reasonable empirical quantiles, as well as preserving temporal non-stationarity). Therefore, we do not report RMSE associated with models that have monthly non-stationary as their interpretation is not helpful. The CRPS does not require the specification of an observation quantile to assess the model performance. Here, we present results for group $A$, model $F_0$, and highlight a handful of the best competing models in Table~\ref{c6:tab_cv_gpd} and discuss our broader findings.

\begin{table}
\centering 
\begin{tabular}{r|lcccc} 
\textbf{Model} & \multicolumn{1}{c}{\textbf{Parameterisation of $\log \sigma_{o}(t,\boldsymbol{s}$)}} & \multicolumn{2}{c}{\textbf{ST-CV}} & \multicolumn{2}{c}{\textbf{$n$-CV}}\\\hline
& & RMSE & CRPS & RMSE & CRPS \\
\hline
$A_0$ & $\beta_0$ & 1.224 & 0.885 & 1.354 & 0.945 \\ 
$A_1$ & $\beta_0+\beta_1\sigma_c(\boldsymbol{s})$ & 1.225 & 0.885 & 1.354 & 0.945 \\ 
$A_2$ & $\beta_0+\beta_1C(\boldsymbol{s})$ & 1.189 & 0.878 & 1.322 & 0.939 \\  
$F_0$ & $\beta_0+\beta_1C(\boldsymbol{s}) + \beta_2M^{I}(t)$ & 1.189 & 0.878 & 1.320 & 0.939 \\ 
\hline
$D_1$  & {$\beta_0 + \beta_1\sigma_c(\boldsymbol{s}) + \beta_2 C(\boldsymbol{s}) +  \beta_3M^{I}(t) + \beta_4M^{I}_{R,m}(t)$} & {-} & {0.815} & {-} & {0.874} \\ 
$F_1$   & {$\beta_0+\beta_1C(\boldsymbol{s}) + \beta_2M^{I}(t) + \beta_3M^{I}_{R,m}(t)$} &{-} & {0.816} & {-} & {0.874} \\ 
$G_1$  & {$\beta_0+\beta_1C(\boldsymbol{s}) + \beta_2M^{I}(t) + \beta_3M^{I}(t)C(\boldsymbol{s}) + \beta_4M^{I}_{R,m}(t)$} & {-} & {0.816} & {-} & {0.873} \\  \hline
\end{tabular}
\caption{Cross-validation metrics for a selection of GPD scale parameter models.}\label{c6:tab_cv_gpd} 
\end{table}

For base models, the incorporation of the climate-model-derived covariate $\sigma_c(\boldsymbol{s})$ in model $A_1$ yields no improvement over the spatially stationary model $A_0$. This is unexpected as this covariate was found to be highly effective for extremely hot summer temperatures \citep{Healy2023} and the climate model was informative for the bulk of the distribution. However, the results in Figure~\ref{fig:blk_qnt_reg_params} show that this finding is consistent with our bulk distribution modelling of $F_B$. In particular, the coefficient of the climate model covariate for quantiles decreases in size as $\tau$ increases, suggesting that the climate model does not capture well the most extremely cold events. In contrast, model $A_2$ shows that the incorporation of the coastal proximity covariate does provide an improved cross-validation score. Model $F_0$ shows that, highly surprisingly, there is no improvement in fit by including the long-term smoothed temperature anomalies over Ireland, i.e., $M^I(t)$. However, as identified in Section~\ref{sec:temp-covariates}, there are complex temporal changes, such as the very cold winter of 2009/10, that may be obfuscating the long-term climate change signal over 1950-2022, so here we additionally investigate the effects of the SCV.

Model $F_1$ is our selected best choice using the criteria, but other models perform well, producing very similar metrics, and so we must discuss our choice. The best fitting models all favour using HadCRUT5 residuals, $M^I_R(t)$, as a covariate for explaining SCV, with the CRPS metric indicating a preference for the monthly covariate $M^I_{R,m}(t)$ rather than the winter averaged $M^I_{R}(t)$. This is somewhat unsurprising as all temperature-related monthly average climatic forcings are encoded within HadCRUT5. Of the other SCV covariates the following also performed well $\{\text{AO}_R, \text{NAO}_R \text{ and, } M_R^A\}$. Each of the best-performing models has coastal proximity as a spatial covariate.  This narrows the selection down to models $D_1$, $F_1$, and $G_1$. Model $D_1$ can be ruled out as it is simply $F_1$ with the additional covariate, $\sigma_c(\boldsymbol{s})$, which was not found to be informative when fitting model $A_1$. We exclude model $G_1$, as the added complexity of the spatio-temporal interaction between $C(\boldsymbol{s})$ and $M^I(t)$ was not justified, since the bootstrap evaluated 95\% confidence interval for its coefficient contained 0. So we chose model $F_1$ for its parsimony, predictive and modelling performance, and ease of interpretation. 
 
Before we move on to examine the features of the model $F_1$ fit, we first consider how the fit of $F_1$ and $F_0$ compare, i.e., the impact of the inclusion of the covariate $M_{R,m}^I(t)$. Clearly, both cross-validation scores improve with this SCV covariate's inclusion, but what is most telling is the change in the estimate of the $\beta_2$ coefficient for $M^I(t)$ in the two models. Specifically, $\hat{\beta}_2$ and associated bootstrap-based 95\% confidence intervals (negated to reflect the data transformation) are $-0.053~ (-0.496, 0.858)$ for $F_0$ and $0.372~ (0.106,0.588)$ for $F_1$. Hence, model $F_0$ is estimating a cooling of extreme cold winter temperature from changes in the scale parameter as the mean temperature is warming, whereas model $F_1$ has the opposite trend, i.e., a warming. So, if we did not account for the jet stream behaviours on the short-term behaviour, we would estimate that extremely cold winter temperatures were getting colder rather than warming, as the better fitting model suggests. There are a few points to help clarify what is going on. Firstly, the confidence intervals for $\hat{\beta}_2$ from model $F_0$ are much wider than for the corresponding estimate using model $F_1$, and they include both zero and the positive $F_1$ point estimate. The reason they are so wide is that the SCV effect is so strong over this window, with the winter of 2009/10 highly influential in producing a negative trend with mean temperatures. By incorporating the zero-mean covariate $M_{R,m}^I(t)$, which is orthogonal to $M^I(t)$,  we can account for winters like 2009/10. This results in a much more plausible long-term trend estimate with increased certainty, i.e., the $\hat{\beta}_2$ confidence interval now excludes zero and negative values. So, the impact of adding in our novel SCV covariate has a profound impact on our scientific conclusions about climate change in cold winters over Ireland.

\begin{figure}[h]
    \centering
    \includegraphics[width = 10cm]{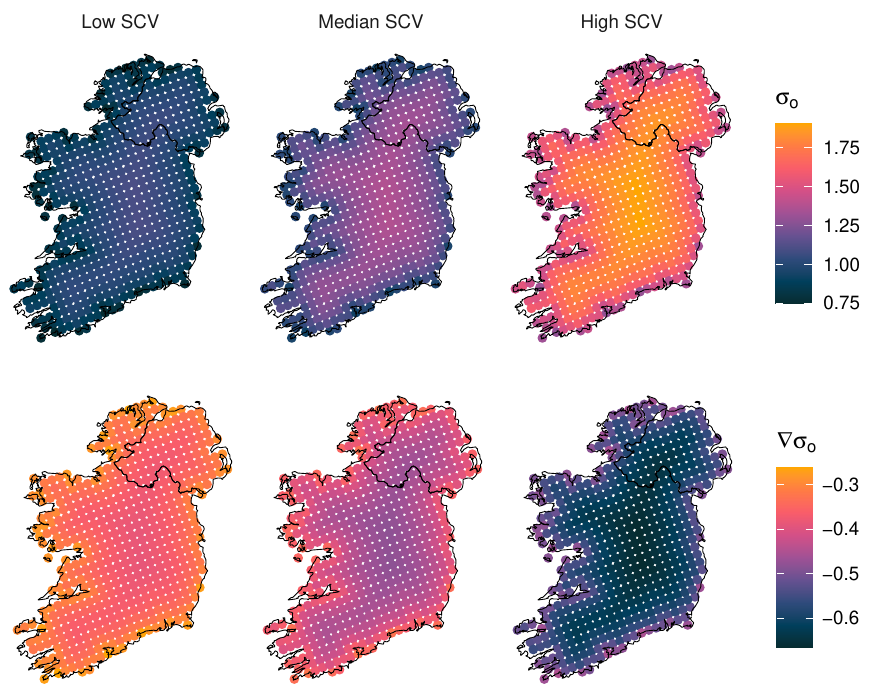}
    \caption{Estimates based on model $F_1$ of the GPD scale parameter in $2022$ (top row) and 
    the change in the scale parameter since $1950$, $\nabla \sigma_o(\boldsymbol{s})$ (bottom row).
    Left to right panels correspond to low, median, and high phases of SCV, respectively.}
    \label{fig:c6_scale_changing}
\end{figure}

For our chosen model, $F_1$, Figure~\ref{fig:c6_scale_changing} plots the estimated scale parameter, $\hat{\sigma}_o(t,\boldsymbol{s})$ in 2022, and its estimated change between 1950 and 2022 of $\nabla \sigma_o(\boldsymbol{s})$, for three different phases of the SCV covariate $M^{I}_{R,m}(t)$. The key feature that this shows is seen from the top row of plots, indicating a substantial change in variability of the coldest temperatures, depending on the state of SCV, with the scale parameter more than doubling as we move from low to high SCV, indicating that the most extreme cold temperatures will occur in a high phase of SCV.  Secondly, the bottom row shows that the scale parameter has reduced over the observed time window for all sites and over all phases of SCV, indicating that extreme cold temperatures have become less likely over the time window, with the largest changes in the high phase of SCV, a drop of $25\%$ relative to 1950. Thirdly, there is evidence that both the scale parameter and its change over time are smallest close to the coast, highlighting the ocean's regulatory effect. 

\begin{figure}[h]
    \centering
    \includegraphics[width = 11cm]{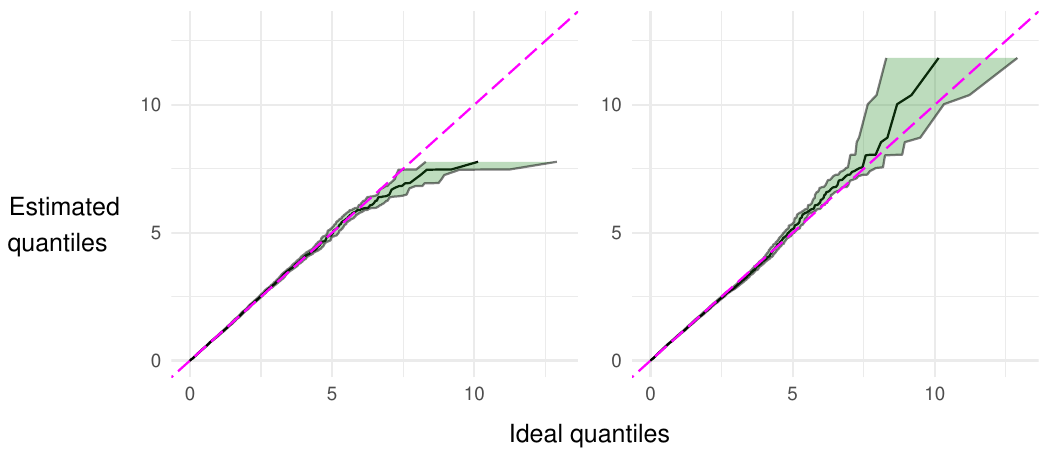}
    \caption[QQ-plot of model with and without climate oscillation covariate.]{Spatial and temporal pooled QQ-plots, on common exponential margins, of models with and without a climate oscillation covariate. Model $F_0$ without SCV covariate (left), final chosen model $F_1$ with SCV covariate (right).
    Shaded regions on each plot give  bootstrap-based pointwise 95\% confidence intervals.}
    \label{fig:c6_qq_plot}
\end{figure}

To assess the absolute quality of the chosen model and the impact of incorporating the best performing SCV covariate $M^{I}_{R,m}(t)$, we present spatial and temporal pooled QQ-plots for models $F_1$ and $F_0$ in  Figure~\ref{fig:c6_qq_plot} along with bootstrap-derived pointwise 95\% confidence intervals. Through these respective fitted models, we transform the data into a common exponential scale under the assumption that the models are correct. The exponential scale accentuates the upper tail of the negated data and highlights the model's performance in capturing the extreme values. Respectively, models $F_1$ (and $F_0$) appear to fit the tail of cold temperatures very well (ineffectively) as the confidence intervals cover (exclude) the line of equality in this region of the distribution. These results reiterate that the incorporation of the SCV effect is essential for explaining variations in extreme cold temperatures over Ireland.  

A further revealing feature from the comparison of models $F_0$ and $F_1$ is that the former has a heavier estimated tail ($\hat\xi_o = 0.011$) than the latter ($\hat\xi_o = -0.079$) and thus overestimates the heaviness of the tail of the distribution giving no finite bound for how cold temperatures can be, see Section~\ref{sec:c6_marg_tail}. In contrast, by accounting for key sources of variation in the extremely cold events, through the SCV covariates, the estimated shape parameter for model $F_1$  gives a more physically realistic finite lower bound for cold temperatures. 

\section{Spatial dependence models}\label{sec:c6_mvd}
\subsection{Nature of spatial extremal dependence}
\label{sec:ADvAI}

To decide on our strategy for extremal dependence modelling of the spatial data, we investigate the nature of the extremal dependence of extreme minimum daily winter temperatures after marginal standardisation. To achieve such standardisation, the marginal distributions are transformed so that they have an identical distribution over covariates to facilitate a fair investigation of the extremal dependence. Specifically, we transform the data to uniform margins using the probability integral transform, and following this, to standard Pareto margins via
\begin{equation}\label{eq:ch6_pareto_margins}
 X_o^P(t, \boldsymbol{s}) = \frac{1}{1-F_o\left(X_o(t, \boldsymbol{s});t, \boldsymbol{s}\right)}, \mbox{ for all }
 \boldsymbol{s}\in \mathcal{S}_o \mbox{ and all }t\in \mathcal{T}_o,
\end{equation}
where $F_o(\cdot; t, \boldsymbol{s})$ is given by formulation~\eqref{eqn:joinedmarginal} with the components of this form given by the covariate models structures of Section~\ref{sec:MargData}. \cite{Coles1999} presents two measures of extremal dependence, $\chi$ and $\bar{\chi}$, to determine the level of bivariate dependence between variables exhibiting asymptotic dependence and asymptotic independence, respectively. In particular, in the spatial context, for the process $\{X_o^P(t, \boldsymbol{s}): (t,\boldsymbol{s})\in \mathcal{T}\times \mathcal{S}\}$, if the process is stationary over $\mathcal{T}$ this measure of dependence for the process at sites $\boldsymbol{s}_i$ and $\boldsymbol{s}_j$ in $\mathcal{S}$ is given by $\chi= \chi^P(\boldsymbol{s}_i, \boldsymbol{s}_j)$ which is defined for all $t\in\mathcal{T}$, by
\begin{align}
\label{eq:chi_lim}
 \chi(\boldsymbol{s}_i, \boldsymbol{s}_j) 
 = \lim_{v \to \infty}\chi_v(\boldsymbol{s}_i, \boldsymbol{s}_j), \mbox{ where }
\chi_v(\boldsymbol{s}_i, \boldsymbol{s}_j)=  \Pr\left\{X_o^P(t,\boldsymbol{s}_j) > v \mid X_o^P(t,\boldsymbol{s}_i)>v\right\}
\end{align}
Here $\chi$ measures the pairwise coefficient of asymptotic dependence between sites $\boldsymbol{s}_i$ and $\boldsymbol{s}_j$. The value of $\chi$ ($0 < \chi\le 1$) determines the degree of asymptotic dependence, with $\chi$ increasing as asymptotic dependence strengthens. The value of $\chi$ is best estimated by using empirical estimates of $\chi_v$ for increasingly large $v$. If $\chi=0$, then process $X_o^P$ is asymptotically independent at this pair of sites, with the nature of the non-limiting dependence best studied by using an empirical estimate $\bar{\chi}_v$ of $\bar{\chi}$.

\begin{figure}[!h]
    \centering
    \includegraphics[width = 13cm]{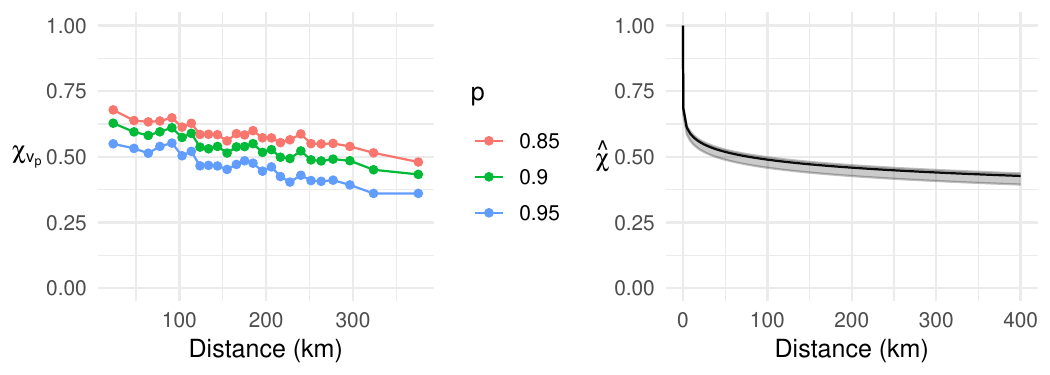}
    \caption{(Left) Empirical estimates of $\chi_{v_p}$ plotted against inter-site distance for standardised observational data, for $p$-th marginal quantile $v_p$ where $p=0.85$, $0.9$ and $0.95$. (Right) $r$-Pareto process parametric estimate of $\chi_o^P$ against inter-site distance $h$ with dark line displaying the value using a Mat\'{e}rn variogram when fitted above an $80\%$ risk threshold, with grey region showing pointwise 95\% confidence intervals, capturing both marginal and extremal dependence bootstrap uncertainties.}
    \label{fig:c6_chi_emp}
\end{figure}

Figure~\ref{fig:c6_chi_emp} presents empirical estimates of $\chi_v$ plotted against the distance between sites for all pairs of sites for the observational data, for three values of $v=v_p$ with $v_p$ the $p$-th marginal quantile with $p=0.85, 0.9, 0.95$. Here, estimates from separate pairs of sites are averaged over sites with similar inter-site distances. Naturally, $\chi_v$ is decreasing with distance, but even for the longest separations between sites over the island, the magnitude of $\chi_v$ is far from zero, and the level of dependence at the furthest ranges is not much reduced relative to those at the shortest. This finding is a consequence of solar insolation being at its lowest in Ireland during winter months, and so factors such as clouds tend to have much lower local impact, leading to large-scale spatial and temporal events \citep{Liou2002}. Although there is a reduction of $\chi_{v_p}$ with increasing $p$, the change is minimal relative to how different the estimates are from zero. It could be viewed that these quantiles are not that extreme, but recall they are quantiles for winter daily minima, meaning that $p=0.95$ corresponds to $4.5$ events on average per year, with any higher quantile leading to highly variable empirical estimates.

\begin{figure}
    \centering
    \includegraphics[width=0.6\linewidth]{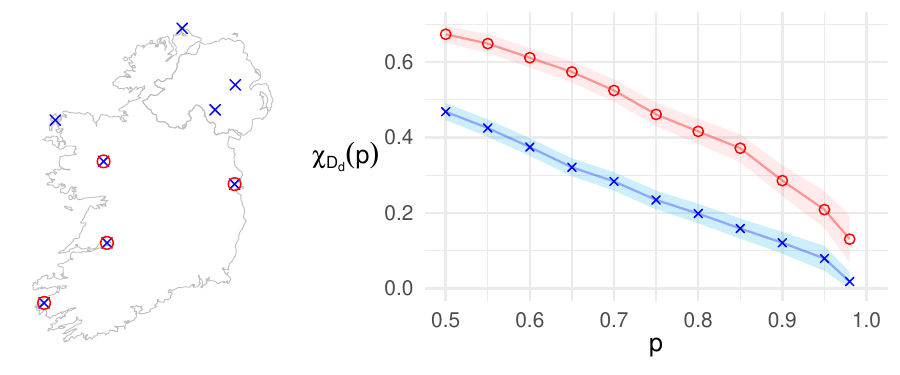}
    \caption{(Left) Map of Ireland with the longest running 4 and 8 station gauges marked as $\textcolor{red}{\circ}$ and $\textcolor{blue}{\times}$ respectively. (Right) Empirical estimates and bootstrap-based 95\% confidence intervals of $\chi_{D_d}(p)$ for the 
    $d=4,8$,  for the stations shown in the left panel, and with corresponding symbols.}
    \label{fig:chi_d}
\end{figure}

{Although Figure~\ref{fig:c6_chi_emp} is broadly supportive of the data being from a process which exhibits asymptotic dependence, formally, this figure only shows asymptotic dependence is consistent with the data (averaged over stations with similar separation distances) for pairs of stations. Following the discussion contribution by \cite{Huser2025Discuss} of \cite{Healy2023}, here we explore the evidence for the temperature process being asymptotically dependent for higher order structures, i.e., dimension $d\ge 3$. To get a sense of this higher order extremal dependence structure of the data, following the approach of \cite{Huser2025Discuss}, we consider the level of asymptotic dependence for a set of $d$ stations, denoted by $D_d$, where $D_d\subset \mathcal{S}_o$. We estimate the extension of $\chi$ to this set of variables, denoted by $\chi_{D_d}(p)$, over a range of percentile levels $p$, where }
\[
\chi_{D_d}(p)=\Pr\{F_o(X_o(t, \boldsymbol{s_j});t, \boldsymbol{s_j})>p: j\in D_d, t\in\mathcal{T}\}/(1-p),
\]
{for extremal events being when $p$ increases towards $1$. As our dataset exhibits large amounts of missing data, this prevents a reliable analysis of the entire network. Instead, we focus on estimating $\chi_{D_d}(p)$ on the four and eight longest running stations, i.e., $d=4$ and $8$. The locations of these two sets of stations are shown in Figure~\ref{fig:chi_d} (left). The empirical estimates of $\chi_{D_d}(p)$, and associated 95\% confidence intervals, for both choices of $D_d$  are given in Figure~\ref{fig:chi_d} (right). These estimates show that the level of extremal dependence among the sites decreases, as  $p$ increases to near 1, corresponding to a marginal event occurring 1.8~days on average per year. Also, as $D_4\subset D_8$ we have $\chi_{D_8}(p) \le \chi_{D_4}(p)$ for all $p\in(0,1)$. The estimates for $d=4$ (and $d=8$) provide non-zero (near zero) levels of asymptotic dependence, respectively, but at the 80\% percentile, there is still evidence of a reasonable level of dependence. These findings are not inconsistent with those of \cite{Huser2025Discuss} for the higher-order dependence structure of extreme hot summer temperatures over Ireland. In common with the modelling assumption made for extremely hot summer temperatures over the island in \cite{Healy2023}, we believe that our exploratory data analysis suggests that the assumption of spatial stationarity and asymptotic dependence in the minimum daily winter temperatures over Ireland is a reasonable first-level approximation, and it is one that we take in the remainder of the paper with a model that explicitly assumes that this property holds.}

A potential weakness in making the assumptions of stationarity is that it would not be too surprising to find that stations on the coast have a weaker extremal dependence with other stations than inter-site distance would imply, and  they may possibly be asymptotically independent of more inland sites. In fact, in the marginal analyses, we found that the coastal covariate $C(\boldsymbol{s})$ was important to our analysis. Neither the \cite{Huser2025Discuss} sub-regional analysis, nor our analyses here, explored this feature. Figure~\ref{fig:chi_d} (right) showed weak asymptotic dependence over the $d=8$ but the left panel of this plot shows that the majority of these sites were coastal as the longest running stations have a bias in their location, since they were typically sited on the coast to reflect the historic needs of offshore weather forecasting rather than for our interests in systematic temperature data analysis. 

\subsection{$r$-Pareto process}\label{sec:c6_rparp}
Based on the evidence detailed in Section~\ref{sec:ADvAI}, we make the assumption that the process is asymptotically dependent over all sites. We follow \cite{Dombry2015}, \cite{DeFondeville2018} and \cite{Healy2023} in modelling the extremal spatial behaviour of the marginally standardised process $X_o^P(t,\boldsymbol{s})$ over $\boldsymbol{s} \in \mathcal{S}$ by an $r$-Pareto process. Key to doing this is to identify which events are extreme by defining a risk function $r$, 
\begin{equation}
r(X_o^P):=r\{X_o^P(\boldsymbol{s}): \boldsymbol{s}\in \mathcal{S}\} \in [0,\infty),
\label{eqn:rFunction}
\end{equation}
which provides an ordering to levels of extremity of spatial observations. Here  $r$ must be homogeneous of order 1, i.e., $r(c\textbf{x}) = cr(\textbf{x})$ for any constant, $c > 0$ and with $\min(\textbf{x}) > 0$. Under weak conditions on the process $X_o^P$, we have that
\begin{equation}\label{ch6:rparetoeq}
\operatorname{Pr}\left[v^{-1}X_o^P(\boldsymbol{s}:\boldsymbol{s}\in\mathcal{S}) \in \cdot \mid r\left\{X_o^P(\boldsymbol{s}:\boldsymbol{s}\in\mathcal{S})\right\}>v\right] \rightarrow \operatorname{Pr}\{Y^P_r(\boldsymbol{s} :\boldsymbol{s}\in\mathcal{S}) \in \cdot\},
\end{equation}
as $v\rightarrow \infty$, where $Y^P_r(\boldsymbol{s}: \boldsymbol{s} \in \mathcal{S})$ is marginally non-degenerate in any margins. Here, $Y^P_r$ is called the $r$-Pareto process. If limit~\eqref{ch6:rparetoeq} is a good approximation for large $v$, then those spatial events with a risk function exceeding $v$ will be well-approximated by an $r$-Pareto process. Crucially, $Y^P_r$ can be decomposed into two independent stochastic components as follows:
\begin{equation}\label{ch6:decomp}
 Y^P_r(\boldsymbol{s}) = RW(\boldsymbol{s})
 \mbox{ for all }\boldsymbol{s} \in \mathcal{S},
\end{equation}
where $R$ is a Pareto(1) distributed random variable and can be interpreted as the risk of the process, and $\{W(\boldsymbol{s}): \boldsymbol{s} \in \mathcal{S}\}$ is a stochastic process which describes the spatial profile of the extreme event, i.e., the proportion of the risk at each site. By construction, we have that $R = r\left(Y^P_r(\boldsymbol{s} :\boldsymbol{s}\in\mathcal{S})\right)$ and $r\left(W(\boldsymbol{s} :\boldsymbol{s}\in\mathcal{S})\right) = 1$. The factorisation of expression~\eqref{ch6:decomp} into a magnitude term $R$ and the spatial profile $W$ of the extremal dependence structure enables the separate inference of the unknown $W$ process from $R$ with a known magnitude distribution. This independence of $R$ and $W$, coupled with a known distribution for $R$, is powerful as it allows extrapolation to events larger than those previously observed.

By the decomposition of the $r$-Pareto process, given by expression~\eqref{ch6:decomp}, the spatial dependence structure of $Y^P_r$ is therefore determined by the properties of the spatial process $W$. Following \citet{DeFondeville2018} and \citet{Palacios2020}, we choose to model $W$ using a log-Gaussian stochastic process that is stationary and isotropic over time and space with a parametric variogram. This choice of a semi-parametric spatial process for $W$ affords great flexibility in the structure of extremal dependence that can be achieved within the class of asymptotic dependent processes. As with \citet{DeFondeville2018}, we take the variogram $\gamma$ to determined by the Matérn variogram family, i.e.,  
\begin{equation}
\label{eqn:mat}
 \gamma(h) = \alpha \left\{1 - (2 \sqrt{\nu} h /\phi)^\nu
  2^{1-\nu}\Gamma(\nu)^{-1} \text{K}_\nu (2 \sqrt{\nu} h/\phi) \right\},
\end{equation}
 for inter-site distance $h\ge 0$, with $K_\nu$ being a modified Bessel function of the second kind, and the parameters $(\alpha, \phi,\nu) \in \mathbb{R}_+^3$, used to determine respectively the variance, range, and smoothness of $W$ \citep{banerjee2014hierarchical}, but which indirectly determine similar dependence features of $X_o(t,\boldsymbol{s})$ over $\boldsymbol{s}$.

Next, we consider the choice of risk function $r$, which, together with a level $v$, combines to identify which spatial events are deemed sufficiently extreme for us to fit the limiting $r$-Pareto process. In particular, $r$ gives the functional form of spatial events and $v$ the minimum size of this functional. If we are interested in any spatial extreme event $A$, then we can estimate its probability using the $r$-Pareto process if $A\subseteq \{X_o^P(\boldsymbol{s}\in\mathcal{S}_o): r\left(X_o^P(\boldsymbol{s}\in\mathcal{S}_o)\right)>v\}$, i.e., at the observational stations, the event, $A$, has a risk, $r$, greater than $v$. The risk function can take any form provided it satisfies the properties set out in formulation~\eqref{eqn:rFunction}. As any homogeneous function can be taken from a mathematical perspective, we need to make a selection based on a principled basis and one that can be determined from observational data. The latter point restricts $r$ to be a function only of the sites $\mathcal{S}_o$ The typical approach that has been proposed is to use the spatial average over the observation sites, i.e., $\boldsymbol{s} \in \mathcal{S}_o$, but other choices have been considered, such as the spatial maximum (i.e., spatial minimum in the original scale for our data, leading to a generalised Pareto process \citep{Buishand2008}), spatial quantiles, or other physically motivated summaries \citep{DeFondeville2018}. Our preference is to take the spatial average for the following set of reasons: (i) the $r$-Pareto process decomposition~\eqref{ch6:decomp} involves independent radial and angular variables, and, in particular restricting $R>u\geqslant v$ in any simulation of such a process ensures the risk of each simulation exceeds $u$; (ii) the choice parallels equivalent threshold choices in multivariate extreme values \citep{Coles1994}; (iii) the choice involves the process at all sites equally; and (iv) the simple adaptation of the risk function for situations with missing data, see below. In considering the choice of $r$, it is important to realise that this function is for data transformed onto Pareto marginal scales, which have very heavy tails and hence the average tends to be dominated by the largest component. So on the original marginal scales, the set of events which exceed the spatial average risk is, in practice, very similar to the events with a risk based on the spatial maxima. So, for non-missing data problems, the choice of risk function is not very important in practice. However, due to the volume of missing data, as described in Section~\ref{sec:StationData}, we need to consider a risk function that is appropriate when any spatial observation can have an arbitrary number of sites with missing data. We follow \cite{Healy2023} in using the average over the observed station data for each $t$, i.e., with the associated risk function being
\begin{equation}
r_t\left(X_o^P(t,\boldsymbol{s}): \boldsymbol{s}\in \mathcal{S}_o\right)
=\frac{\sum_{i=1}^{|\mathcal{S}_o|} X_o^P(t,\boldsymbol{s}_i)I_o(t, \boldsymbol{s}_i)}{\sum_{i=1}^{|\mathcal{S}_o|} I_o(t, \boldsymbol{s}_i)}, 
\end{equation}\label{eq:risk_functional} where $I_o(t, \boldsymbol{s}_i)$ is the indicator variable for whether $X_o(t,\boldsymbol{s}_i)$ is observed or not. We make the assumption that the limit in  expression~\eqref{ch6:rparetoeq} holds exactly above a given value of $v$, and take this $v$ to be the 80th percentile of $\{r_t: t\in \mathcal{T}_o\}$.

Using the software of \citet{DeFondeville2018} we found difficulty in estimating the parameters $(\phi,\nu)$ using maximum likelihood inference: for the latter we followed the strategy of \cite{ryan2017} and restricted $\nu$ to a grid of 25 values between 0.01 and 0.8, giving $\hat{\nu}=0.1$; with $\hat{\phi}=142$km. We found a reliable inference for $\alpha$, with $\hat\alpha$ and its bootstrapped $95\%$ intervals found to be $2.63~(2.53, 3.00)$. Figure~\ref{fig:c6_chi_emp} shows the parametric estimate of $\chi$ from using the fitted $r$-Pareto process with Matérn variogram as well as the associated pointwise 95\% confidence intervals based on $200$ spatio-temporal bootstrap samples. The comparison with the empirical estimates over a range of high thresholds shows a good fit to the data, suggesting that the $r$-Pareto process is capturing the spatially varying level of asymptotic dependence well.

\section{Results}\label{sec:c6_results}

\subsection{Marginal return level results}\label{c6:sec_marg_rl}
We present a range of marginal summaries of how daily minimum winter temperature extreme events in Ireland are changing over the period 1950--2022, and in particular how these features change in relation to the three phases of SCV. i.e., the covariate summarises $M^{(0.1, I)}_{R,m}(t), M^{(0.5, I)}_{R,m}(t)$, and $M^{(0.9, I)}_{R,m}(t)$. We also investigate the sensitivity of our return level estimates to the threshold choice made in Section~\ref{sec:c6_threshold_selection}. These results combine the effects of the different covariate relationships on each aspect of the marginal model, i.e., the threshold function, the rate of exceedances of that threshold, and the GPD parameters.

\begin{figure}[h]
    \centering
    \includegraphics[width = 8.5cm]{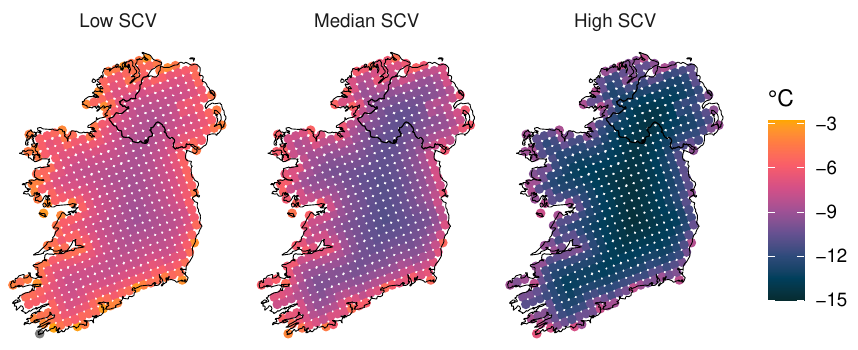}
    \includegraphics[width = 8.5cm]{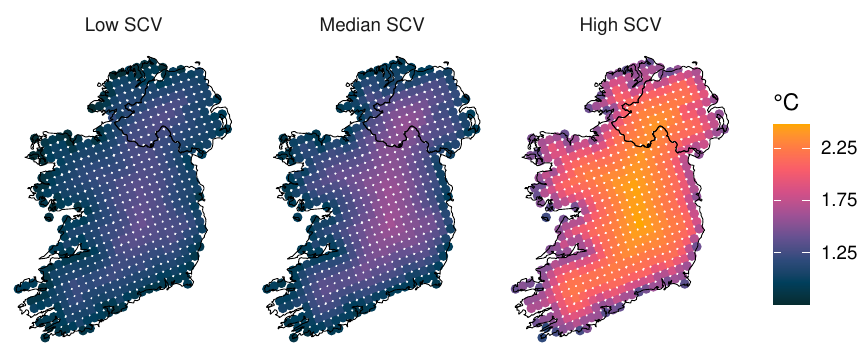}  \caption{(Left) Estimated 100-year marginal return level in 2022 and (Right) half  the width of the 95\% confidence interval of 100-year return level in 2022: shown for (left-right) low, median, and high phases of SCV.}
    \label{fig:100_rl_est_m22}
\end{figure}

First, consider how return levels vary with the novel covariate SCV. Figure~\ref{fig:100_rl_est_m22} presents estimates of the 100-year return level in the context of the year $2022$ for the three different phases of SCV based on our selected best-fitting model $F_1$. These conditional 100-year return levels (i.e., an event with occurrence probability $1/(100\times90.25)$ on a given day) are presented for each phase of SCV. Figure~\ref{fig:100_rl_est_m22} also shows the half the width of the 95\% confidence interval for these return level estimates, with these values being derived from $200$ spatio-temporal bootstrap samples, see Section~\ref{sec:c6_bts}. Across the low to high phases of SCV, these results show that there is a clear and statistically significant decrease in extreme cold winter temperatures with greater uncertainty in estimates of the coldest events. In particular, return levels are found to be in the ranges $-8.9^\circ$C to $-2.7^\circ$C, $-10.9^\circ$C to $-3.9^\circ$C, and $-14.6^\circ$C to $-6.8^\circ$C over Ireland for the respective low, medium and high SCV covariates. Suggested rewording: Inland sites experience more extreme cold and higher uncertainty in their return levels than coastal ones, irrespective of the phase of SCV.

\begin{figure}[h]
    \centering
    \includegraphics[width = 8.7cm]{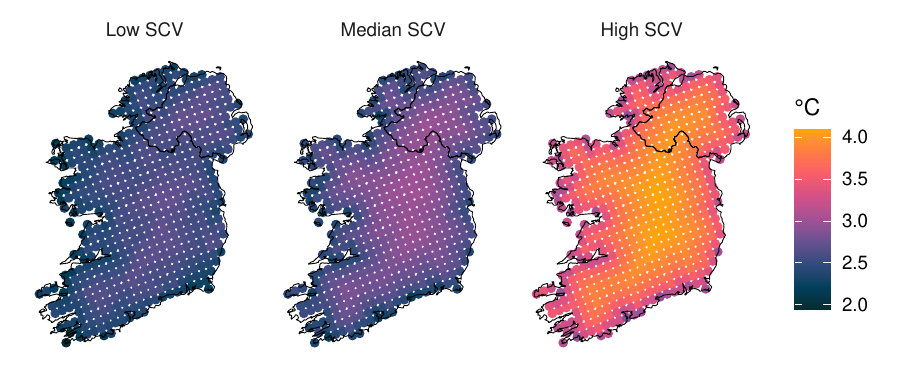}
    \includegraphics[width = 8.7cm]{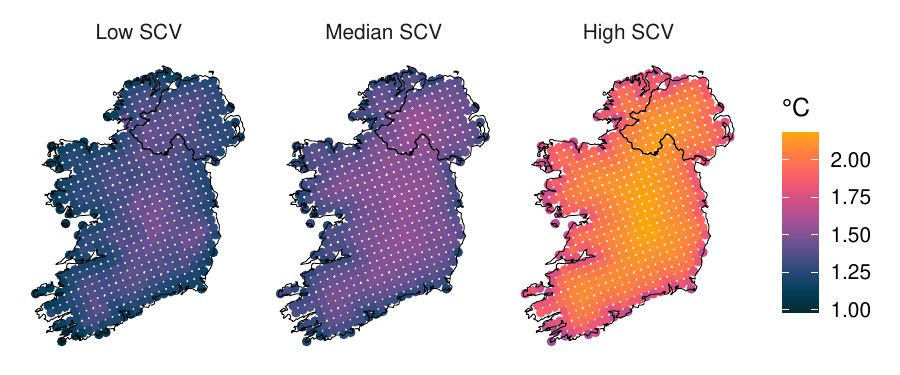}  \caption{(Left) Estimated change in 100-year marginal return level over the period 1950--2022 for (left-right) low, median, and high phases  of SCV, with a positive change showing a warming of the coldest winter temperatures and (Right) half the width of the 95\% confidence interval change in 100-year return level according to model $F_1$ shown for (left-right) low, median, and high phases of SCV.}
    \label{fig:change_100_yr_rl}
\end{figure}

Figure~\ref{fig:change_100_yr_rl} shows similar information to Figure~\ref{fig:100_rl_est_m22}, but here for the estimated change in the 100-year level from 1950 through to 2022, with positive values in the differences, which occur in all panels, representing a warming of the coldest winter temperatures over this time period. The warming of these extremely cold winter temperatures changes substantially with the SCV phase, with the largest change being in the high SCV phase, with an estimated increase of $2.9$--$4^\circ$C over the island. This suggests that climatic conditions that would typically result in very cold temperatures in Ireland are now resulting in much warmer temperatures. This aligns with the findings of \cite{CHRISTIANSEN2018}, which suggest that the winter of 2009/10 was much milder than would have been historically expected given the climatic conditions at the time. The lowest estimated warming, in the low phase of SCV, corresponds to an increase of $1.9$--$2.7^\circ$C over the island. However, even that increase is substantially larger than the approx $0.8$--$1^\circ$C change of mean temperature levels over the island over the same period, i.e., the change in $M^I(t)$ shown in Figure~\ref{fig:c6_mean_temp_change}. Half widths of the 95\% confidence intervals of the differences in return levels also increase with SCV phase so the largest warming estimates are also the most uncertain, so some caution is required in over-interpreting these changes, but it does suggest that the effect of the phase of SCV on the warming of extremely cold winter temperatures needs to be much more widely investigated globally. These findings strongly reiterate that climate change is more radically affecting extreme temperatures than mean temperatures. Interestingly, the warming found here for winter 100-year cold temperatures is greater than the corresponding increase for summer 100-year  hot temperatures over a similar period (1942--2020), with \cite{Healy2023} estimating that to be $1.2$--$2.2^\circ$C. This indicates that winter extreme minimum temperatures are warming faster than summer extreme maximum temperatures, corroborating Ireland's reflection of global trends \citep{IPCC2021}. Finally, irrespective of SCV phase, Figure~\ref{fig:change_100_yr_rl} shows that the largest warming in winter extreme cold temperatures occurs inland.

\begin{figure}[h]
    \centering
    \includegraphics[width = 8cm]{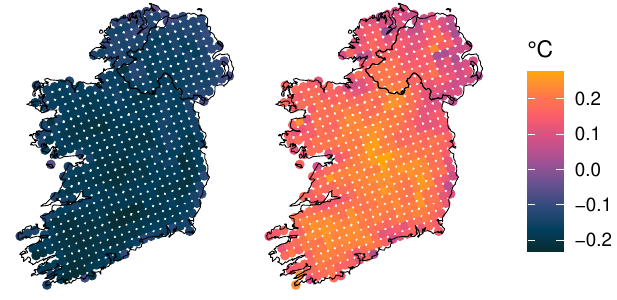}
    \caption[The difference in the 100-year return level during the median phase of SCV in 2022 for a higher and lower threshold.]{The difference in the 100-year level during the median phase of SCV in 2022 between model $F_1$ fit to exceedances above the chosen threshold of $u^{(0.96)}(\boldsymbol{s})$ and to exceedances above both a lower threshold of $u^{(0.95)}(\boldsymbol{s})$ (left) and higher threshold of $u^{(0.97)}(\boldsymbol{s})$ (right).}
    \label{fig:c6_diff_in_rl_diff_thresh}
\end{figure}

Figure~\ref{fig:c6_diff_in_rl_diff_thresh} presents the difference between the estimated 100-year levels for the median SCV covariate, derived using our chosen threshold, $u^{(\tau)}(\boldsymbol{s})$ for percentile $\tau=0.96$, with corresponding estimates for $\tau = 0.95$ and  $0.97$. These estimates show that the return level increases vary slightly with increasing $\tau$, with the differences much smaller than the half-width of the 95\% confidence interval of the differences in return levels shown in  Figure~\ref{fig:change_100_yr_rl}. This shows that there is much less sensitivity to a sensible range of threshold choices relative to the sampling-based uncertainty for a given threshold, indicating that our specific choice of threshold is not an important factor in determining our findings. 

\subsection{Extreme spatial event results}\label{c6:res_spatial_res}

We now summarise temporal non-stationarities over the period 1950--2022 in the spatial context, showing how the estimated time evolution of the marginal distributions, $F_o(\cdot; t, \boldsymbol{s})$, over $\boldsymbol{s}\in \mathcal{S}$, affects the characteristics of winter extreme cold temperatures over Ireland. To derive the various characteristics of spatial extreme events, we first generate a database of spatial extreme simulation events from the $r$-Pareto process, on Pareto margins, and then back transform to the data scale using the inverse of expression~\eqref{eq:ch6_pareto_margins}. Secondly, we present empirically estimated spatial risk measures using these simulations. The specific details of the first step are straightforward since our model has that the extremal spatial dependence structure is invariant to time, so we can generate identical $r$-Pareto process spatial events and then transform these to the original marginals corresponding to any specific time, and any phase of SCV using our selected marginal model. The $r$-Pareto process simulations are generated using the R package \texttt{mvPot} \citep{DeFondeville2021}, which exploits the decomposition in~\eqref{ch6:decomp}. Point estimates of the spatial risk summaries are based on $100,000$ simulations for each phase of SCV. Uncertainties are based on $10,000$ simulations for each $r$-Pareto process, transformed to the original marginal distributions, fitted with marginal and spatial dependence models, to each of the $200$ spatio-temporal bootstrap datasets, totalling an additional two million simulations for each phase of SCV. 

\begin{figure}[h]
    \centering
    \includegraphics[width = 14cm]{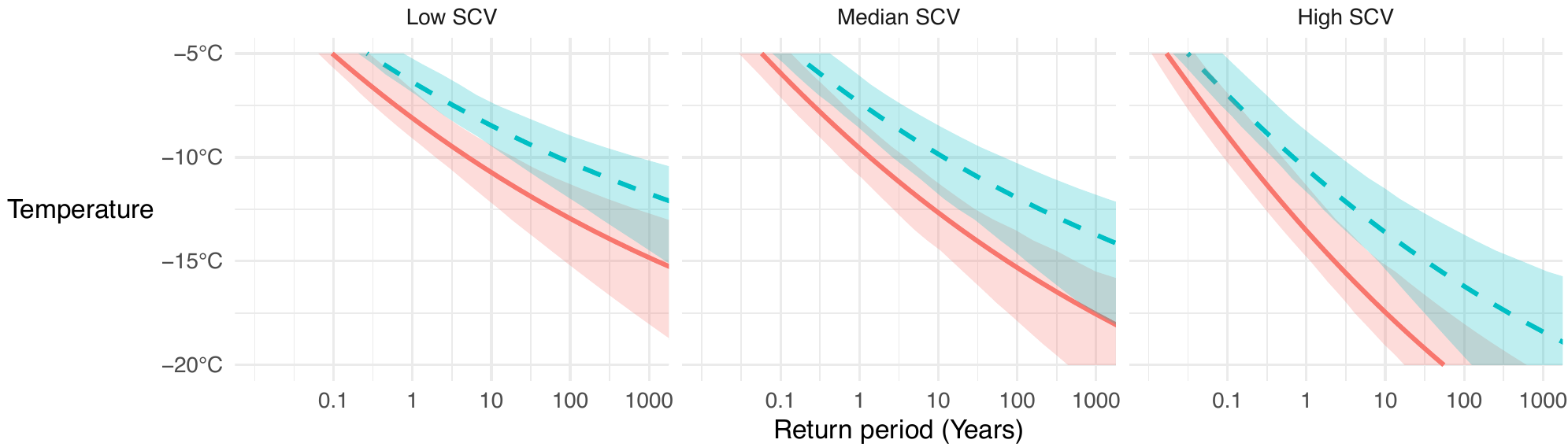}
    \caption[Return period of an extreme temperature anywhere on the Irish observational grid during each level of SCV.]{Return period (on the $x$-axis) of the event $A_{t,\mathcal{S}}(T)$, corresponding to an extreme temperature exceeding $T^{\circ}$C occurs somewhere on the Irish station network $\mathcal{S}_o$: for low, median, and high SCV phases (left-right). Blue dashed (solid orange) lines correspond to $t=2022$ ($1950$) respectively. Shaded regions show corresponding pointwise 95\% confidence intervals.}
    \label{fig:c6_spatial_rl}
\end{figure}

We are interested in making inferences about spatial events of the observational process that are more extreme than a critical temperature of $T^\circ$C somewhere over the observed stations in Ireland at time $t$. We denote these events by $A_{t,\mathcal{S}_o}(T)$, which we can express mathematically in terms of events for the $r$-Pareto process of the form
\begin{align}\label{eqn:ch5ExtremeSet2}
 A_{t,\mathcal{S}_o}(T)&=\left\{X^P_o(t,\boldsymbol{s}), \boldsymbol{s}\in \mathcal{S}_o: \exists ~\boldsymbol{s}_*\in \mathcal{S}_o \mbox{ with }X^P_o(t,\boldsymbol{s}_*)>T^P(t, \boldsymbol{s}_*)\right\},
\end{align}
with $T^{P}(t, \boldsymbol{s})$ being the mapping of $T^{\circ}$C through the transformation~\eqref{eq:ch6_pareto_margins} at time $t$ and for site $\boldsymbol{s}$. Recall that we are modelling negated minimum daily temperatures, so the set  $A_{t,\mathcal{S}}(T)$ in \eqref{eqn:ch5ExtremeSet2} corresponds to extremely cold temperatures. To estimate $\Pr\left\{A_{t,\mathcal{S}}(T) \right\}$ brute force Monte Carlo is highly inefficient, and we follow the importance sampling procedure of \cite{Healy2023}, which has a scaling factor $b_{T(t)}$, which we explain below. Mathematical justification for our estimate $\widehat{\Pr}_{imp}\left\{ A_{t,\mathcal{S}}(T)\right\}$ is given by \cite{Healy2023}. Specifically,
\begin{equation} 
\widehat{\Pr}_{imp}\left\{ A_{t,\mathcal{S}_o}(T)\right\}
 = 
 \frac{1}{b_{T(t)}mL}\sum_{i = 1}^m
 \sum_{j = 1}^L
 I\left\{\exists \boldsymbol{s_0} \in \mathcal{S}_o: r^P_j b_{T(t)}
\frac{y^P_i(\boldsymbol{s_0})}{r_i} > T^P(t, \boldsymbol{s_0}) \right\},
 \label{eqn:ch_4_imp}
\end{equation}
where $m=10,000$ is the number of simulated fields, with the $i$-th field denoted by $y^P_i$, each with corresponding cost $r_i$, and $\{r^P_j; j = 1, \dots, L\}$ are realisations of independent and identically distributed unit Pareto variables, and $L$ is the number of importance sampled events per simulated field, which we take to be $300$. {The optimal value of the scaling factor $b_{T(t)}$ is selected to ensure that at least one of the importance sampled realisations, per simulated field, falls in the set  $A_{t, \mathcal{S}_o}(T)$ of interest, i.e., achieves at least the temperature $T^{\circ}C$ somewhere in $\mathcal{S}_o$ in year $t$. \cite{Healy2023} derives that $b_{T(t)} = \min_{\boldsymbol{s}\in \mathcal{S}_o} \left\{ T^P(t, \boldsymbol{s}) / \omega_{(m)}(\boldsymbol{s}) \right\}$, where $\omega_{(m)}(\boldsymbol{s})$ is the largest (rescaled) contribution at site $\boldsymbol{s}$ among the $m$ independent simulations of the spatial field $W(\boldsymbol{s})$ (as defined in expression~\eqref{ch6:decomp}), i.e., $\omega_{(m)}(\boldsymbol{s}) = \max_{i=1, \dots, m}\left\{{y_i^P(\boldsymbol{s})}/{r_i}\right\}.$ Estimates, and associated pointwise 95\% bootstrap confidence intervals of $\Pr\left\{A_{t,\mathcal{S}}(T) \right\}$ for a range of extremal temperatures $T \in [-20, -5]^\circ C$ for the years 1950 and 2022 and for different phases of SCV are given in Figure~\ref{fig:c6_spatial_rl}. These estimates, where probability is presented on a return period scale, show that the changes in return period for the occurrence over the island of any critical cold temperature $T^{\circ} C$ vary as much, or more, over SCV than across the time window.}

There has been very limited previous analysis of the spatial behaviour of extremely cold events over Ireland. One exception is an empirical study
of \cite{Rohan1986}, who stated that a temperature of $-10^\circ$C is an extremely cold temperature in Ireland and is in the coldest $1\%$  of daily minima temperatures.
Our model-based estimates in Figure~\ref{fig:c6_spatial_rl} enable us to provide a more detailed investigation into the conclusion of \cite{Rohan1986}. According to our model, over the 1950-2022 period, the return period of a spatial event $A_{t,\mathcal{S}_o}(-10)$ has increased from a 1 in 0.2-year to a 1 in 0.7-year event in high phases of SCV, from a 1 in 1.4-year to a 1 in 12-year event for median phase, and a 1 in 5-year to a 1 in 70-year event for the low phase. As Figure~\ref{fig:c6_spatial_rl} illustrates, our model enables us to consider the return period of events of any level of extremity, 
i.e., well beyond the potential of empirical evidence. For example, if we consider the coldest temperature in the Republic of Ireland during the study period, observed in the winter of 2009/10, which was $-17.5^\circ$C on the 25th of December 2010 in Co.~Mayo, the spatial event $A_{t,\mathcal{S}_o}(-17.5)$ has changed over the period 1950 to 2022 from approximately a 1 in 10-year to a 1 in 370-year event during a high phase of SCV. Regarding the coldest temperature in recorded history in Ireland of $-19.4^\circ$C recorded at Omagh, Co.~Tyrone on the 23rd of January 1881 \citep{Hickey2011}, the spatial event $A_{t,\mathcal{S}_o}(-19.4)$ correspondingly changes from being a 1 in 38-year  to a 1 in 3,700-year event for the high phase of SCV. Note that the pointwise 95\% confidence intervals, given in Figure~\ref{fig:c6_spatial_rl}, become increasingly wide as $T$ decreases, and that this coldest observed temperature is not in our dataset, which starts in 1950. So the precise values of the estimated return periods should be used with caution; the fact that they are changing so substantially should not be taken lightly, as they have major implications for society. 

\begin{figure}[h]
    \centering
    \includegraphics[width = 12.5cm]{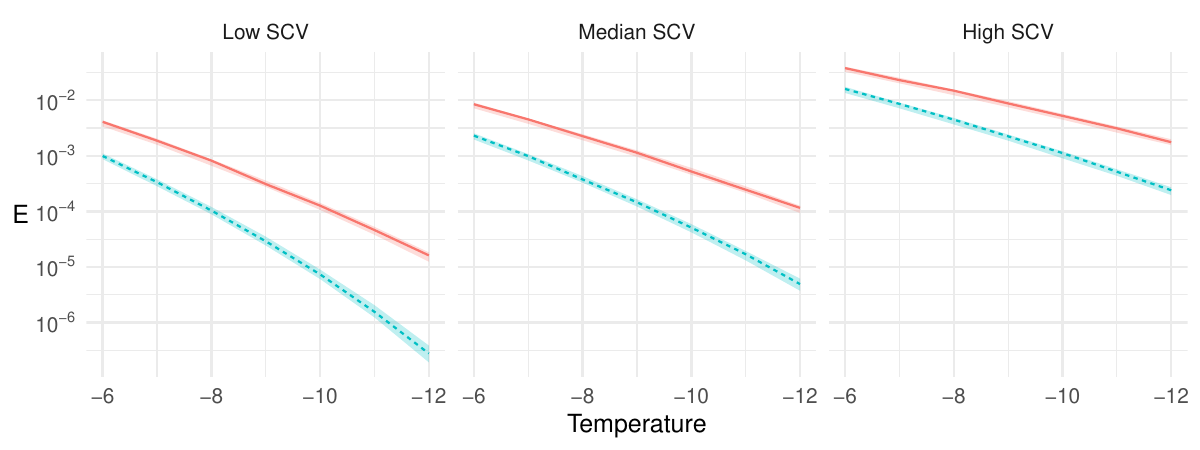}
    \caption{Estimates of the expected proportion, 
    $E(C_T; t, \mathcal{S}_o)$, of Ireland colder than $T^{\circ}$C for a range of $T$ according to model $F_1$ during each level of SCV for low, median, and high levels of SCV (left to right, respectively) in 1950 (solid, orange line) and 2022 (dashed, blue line). Confidence intervals are based on $10,000$ simulations for each 200 bootstrap sample dataset and each level of SCV.}
    \label{fig:c6_prop_ex}
\end{figure}

Finally, in Figure~\ref{fig:c6_prop_ex} we present estimates of the expected value of the proportion, $C_T$, of the observation station network over Ireland which are colder than $T^\circ$C on any given day, conditionally on being in each of the three SCV phases and in the years 1950 and 2022. Specifically, the expected proportion of stations is denoted by $E(C_T; t, \mathcal{S}_o)$  in year $t$, where
\begin{align}
E(C_T; t, \mathcal{S}_o) = \mbox{E}\left( \frac{1}{|\mathcal{S}_o|}\sum_{i=1}^{|\mathcal{S}_o|}
I\{X_o(t,\boldsymbol{s}_i) > T\} \right)
=\frac{1}{|\mathcal{S}_o|}\sum_{i=1}^{|\mathcal{S}_o|}
\Pr(X_o(t,\boldsymbol{s}_i) > T),
\end{align}
where $I(B)$ is the indicator function of event $B$; recall the data are negated. For this functional, the spatial dependence does not influence the point estimates, but it does affect their associated confidence intervals. 
Over the study period, Figure~\ref{fig:c6_prop_ex} shows that the relative change in expected proportion of the observation station network exceeding an extreme temperature $T^\circ$C, decreases with increasingly extreme cold temperatures. It also shows that for $T=-10^{\circ}$C, estimates of $E(C_T; t, \mathcal{S}_o)$ have decreased by a factor of 17.2, 10.2, and 4.68 in phases of low, median, and high SCV. Further analysis (presented in the Supplementary Material Section~5) investigates spatial extremal behaviour through a data-scale measure of extremal dependence, which combines temporal changes in the marginal distributions with the estimated extremal dependence structure. These results indicate that, while joint cold extremes have become less frequent over time, their spatial footprint has also reduced since 1950.

\section{Spatial temporal bootstrapping procedure}\label{sec:c6_bts}

{We rely on bootstrapping to characterise and propagate the uncertainty in our modelling assumptions throughout the entire analysis, from the marginal bulk and tail models to the $r$-Pareto process inference and simulations. We found that the bootstrapping procedure used by \cite{Healy2023} was not applicable to our winter dataset for two reasons. Firstly, our dataset of extremely cold winter events presents substantially stronger temporal dependence, with some extreme cold events lasting over a month to possibly a winter season, which their bootstrapping algorithm didn't need to account for. Secondly, we found that temporal trends in cold winter extremes were much more variable and more complex in terms of dependence on climate indicators than for hot summer extremes due to slow-moving atmospheric conditions (such as those caused by prolonged negative phases of the NAO or AO). When using the standard block bootstrap method, as formulated by \cite{Healy2023}, we found that periods such as winter 2009/10, which contain a disproportionately large portion of extremal observations as seen in Figure~\ref{fig:hadcrut_irel_v_glob}, greatly affected extremal inference on that bootstrapped dataset. This was because, in that bootstrap method, the winter was incorrectly separated into a number of presumed independent blocks (of equal length)  and so it was more likely to be resampled.}

{We use a novel customised bootstrapping algorithm designed to preserve the complex spatial and short-term temporal dependence in the process, while also matching missing data patterns in the observational data. There are five steps to our method: (i) identify in time the peak of an event which is extremely cold at some location in Ireland; (ii) around the peak times, partition all the observed data to construct spatio-temporal blocks of data which capture the full duration (onset, peak, and decay) of a cold temperature event and its spatial footprint over Ireland; (iii) transform each of the temperatures over time and space in each block to a uniform distribution scale to make them time and covariate independent; (iv) resample the blocks of the partition independently from the set of blocks of transformed data to construct a series of the length of the observed data, with the resampling ensuring a comparability with durations, missing data patterns, and dependence structure for the observed data at that time; and (v) the blocks of transformed data are back-transformed to the temperature scale using the covariates specific to the new time period they have been moved to. Details of, and justification for, each of these steps are set out below.}

\begin{figure}
   \centering
    \includegraphics[width=\linewidth]{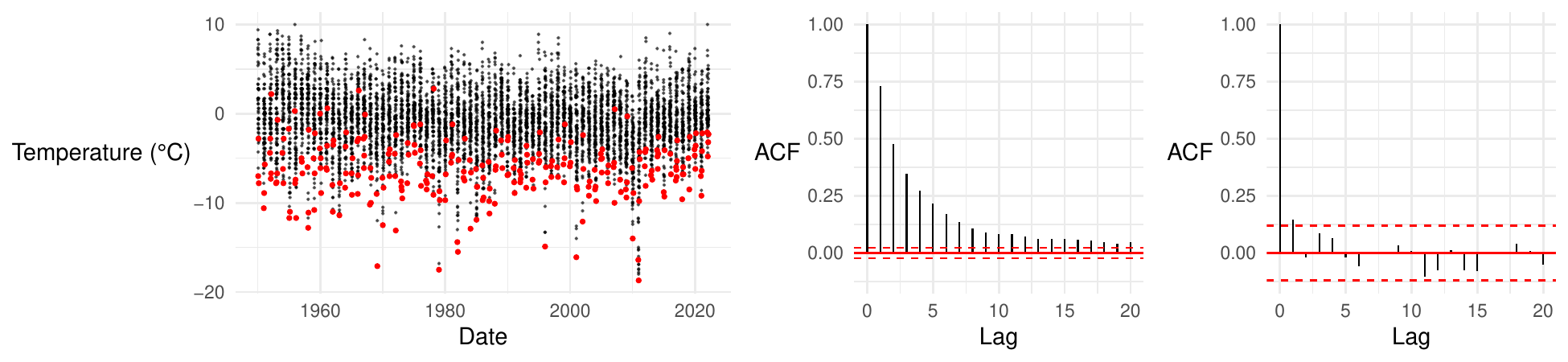}
    \caption{(Left) Time series of the minimum temperature observed across all sites in Ireland for each day between 1950--2022, with declustered event peak data highlighted in red.
   (Centre and Right) Autocorrelation function of the minimum daily temperature recorded anywhere in Ireland and for the declustered event peak data, respectively.}
    \label{fig:ts_min_daily}
\end{figure}

{To achieve step~(i), we create the univariate time series of the daily minimum temperature over all stations in Ireland,  with this series shown in Figure~\ref{fig:ts_min_daily} (left). The cold winter of 2009/10 is clearly seen on this plot to produce a number of the coldest days over 1950-22 across Ireland. Figure~\ref{fig:ts_min_daily} (centre) illustrates that the constructed time series exhibits temporal dependence,  with non-negligible autocorrelation persisting up to 10-15 days. To identify the peak time of an extremely cold event, we iterate sequentially through the time series, selecting local minima. If two local minima fall within 15 days of each other, we keep the colder temperature value and continue until no two local minima are within 15 days of each other, with this value of the lag being motivated by our conclusions about approximate independence of the series based on Figure~\ref{fig:ts_min_daily} (centre). The declustered event peak data are highlighted in red in Figure~\ref{fig:ts_min_daily} (left), with the autocorrelation plot for these, given in Figure~\ref{fig:ts_min_daily} (right), showing that these peak values are approximately independent in time. The required times of the peaks of distinct cold events are therefore set to be the dates of these declustered event peaks. As a result of this process, we find 269 peak events over the 73 years of data, i.e., 3.7 peaks per year, with three of the peaks being identified in the winter of 2009/10 as the continuous cold period in that winter lasted longer than other winters.}

{In step~(ii), the series of declustered event peaks in Figure~\ref{fig:ts_min_daily} (left) are used to deterministically partition all of the spatio-temporal daily minimum temperature data into distinct blocks, which have the property that each block covers at least one cold temperature data value somewhere over Ireland. Critically, this avoids the shortcomings of using the arbitrary fixed-length blocks approach of \cite{Healy2023}, which we have identified fails to work well. For each peak time, the associated block extends halfway backwards (forwards) in time to the previous (next) peak, respectively. Hence, blocks cover the full duration and spatial footprint of the event; they have random durations, and they have random numbers of spatial sites with data, as this number will depend on the missing data pattern for that time period. Although the spatial minimum value depends on the number of stations with observed data at that time, we do not expect the number of stations to vary much within a block, so missing data should not cause any serious issues in the block identification. The blocks are assumed to be independent in time and capture the short-range temporal dependence in the data, which reflects the potential for long-lasting cold periods in winter.} 

{For the $j$th peak time, $t_j$, we define the block start and end times of the associated block by $t^L_j$ and $t^U_j$, with $t^L_j<t_j<t^U_j$,
then the $j$ block ($1\le j\le 269$) constitutes the values}
\[
P_j=\{X_o(t,\boldsymbol{s}): t^L_j\le t\le t^U_j, \boldsymbol{s} \in \mathcal{S}_o\},
\]
{and $t_{j-1}^U= t^L_j-1, t_{j+1}^L= t^L_U+1$, with $t^L_1$ and $t_{269}^U$ being the first and last dates in the dataset and the values with missing data being ignored. In step~(iii), we transform all variables over space and time to be uniformly distributed, i.e., the values in $P_j$, for $j=1, \ldots, 269$, to }
\[
P_{U,j}=\{X^U_o(t,\boldsymbol{s}): t^L_j \le t \le t^U_j, \boldsymbol{s} \in \mathcal{S}_o\},
\]
{where $X^U_o(t, \boldsymbol{s}) = F_{o}\left\{ X_o(t, \boldsymbol{s}); t,\boldsymbol{s}\right\}$, with $F_{o}$ given by our estimate of distribution function~\eqref{eqn:joinedmarginal} at time $t$ and site $\boldsymbol{s}$. This transformation makes the values in each block independent of time and covariates, but retains both spatial and temporal dependence of the data around peak cold events.}

{For our bootstrap, we resample blocks from the partition. If the marginal model was correct, the dependence structure was invariant to covariates, and there was limited missing data, then blocks could be sampled entirely at random. But due to the substantial volume of missing data, \cite{Healy2023} found that some restrictions were needed to ensure that, after bootstrapping the partition blocks, the missing pattern in the resulting bootstrapped sample was close to that of the observed data. We first applied the approach of \cite{Healy2023} but found that this led to relatively few suitable choices available to resample, leading to little variation in the bootstrapped datasets. }

{To overcome the limitations of the previous method, we experimented with a relaxed resampling algorithm for spatial missing data pattern matching until we found a method that worked well. Specifically, we accept a resampled block if it has data for a high proportion of the sites in the block it was replacing, with at least 90\% proportion being found to be reasonable for our needs. In order to preserve the pattern of missing data in a block relative to the block it is replacing, we discarded observations from sites in the resampled block that are not in the block that it was replacing. As block sizes are random, the bootstrapping must ensure the same sequencing of the lengths of blocks are retained in the bootstrap sample so that the overall lengths of samples match up to the observed data for each winter. So, if the resampled block was of a longer duration than the block being replaced, we randomly choose a contiguous subset of the resampled block with the required length. To address potential non-stationarity in the spatial and temporal dependence structure, bootstrapped blocks were not moved by more than 10 years from their original occurrence times.}

{Under the chosen marginal model, the values in the accepted bootstrapped blocks are then transformed back to the data scale to give the full bootstrap sample. To give a sufficient illustration of what this involves, without getting into the full complexity of partitions of different lengths and the associated notation that requires, consider here a partitioned block starting at time $t_j^L$ being bootstrapped to a start time of another block, numbered $j^b$ in the partition, with exactly the same length. Then the back-transformed bootstrapped of block $P_j$ gives the bootstrapped block $P^b_{j^b}$, starting at time $t_{j^b}^L$, of }
\[
P^b_{j^b}=\left\{X^b_o(t, \boldsymbol{s}) = F_{o}^{-1}\left[ X^{U,b}_o(t-t_{j^b}^L+t_j^L, \boldsymbol{s});
t,\boldsymbol{s}\right]:
t^L_{j^b}\le t \le t^U_{j^b}\right\}.
\]
{Thus, block $j$ on uniform marginals is transformed back to the original temperature marginal scale using the fitted marginal model that corresponds to the time window for block $j^b$. Hence, this simulated block has the spatial and temporal empirical dependence of partition block $j$ and the marginal behaviour of partition block $j^b$.}

\section{Conclusion \& discussion}\label{sec:c6_conclusion}

The aim of our research was to identify and characterise non-stationarities of extremely cold daily winter temperatures in Ireland. Given the complex and variable nature of extreme minimum winter temperatures, we emphasise the importance of considering the climatic context (i.e., the atmospheric and oceanic patterns) in which they occur. In particular, we explore the relative importance of both long-term and shorter-scale variations on climate over the period 1950-2022. Specifically, we investigate how shifts in the jet stream's behaviour can lead to the increased occurrence of extremely cold winter temperatures. We have presented some novel candidate approaches to account for large variations in cold extremes during different phases of short-term, but large spatial-scale, climatic variability. These short-term variations are observed over time scales ranging from about a month to an entire winter season. From a range of potential covariates for this behaviour, we find that using the HadCRUT5 dataset, which combines land and sea surface mean monthly temperature anomalies, best captures and represents all climatic processes driving extremely cold temperatures.  We have presented methods for characterising and visualising climate risk associated with different phases of the short-term climatic variability, in particular, by using the HadCRUT5-based covariate, $H_{R, m}^I(t)$, defined in Section~\ref{sec:c6_short_term_covariates}. We have developed a novel bootstrapping approach (used to account for uncertainty throughout) to accommodate the stronger temporal dependence and complex variability observed in extremely cold winter events, as compared to hot summer events we have previously studied in \cite{Healy2023}, where we used a more standard bootstrap procedure.

Our study reveals that, across Ireland, there has been a decrease in the frequency and intensity of extremely cold temperatures since 1950. Furthermore, the rate of warming of extreme minimum temperatures is substantially greater than that of maximum summer temperatures \citep{Healy2023}, or mean temperatures, over Ireland. The rate of these changes varies, with the smallest (largest) changes in coastal (inland) regions, respectively. We identify a profound effect on the long-term trend estimation from the short-term climatic variability, as this factor explains a greater proportion of the variation in extreme cold winter temperature over Ireland than is given by the long-term trend. In particular, we found that if this covariate was ignored, the trend would falsely be estimated to imply cold winter temperatures were getting colder, not warmer. We found that from 1950 to 2022, the occurrence rates of extremely cold winter temperatures have decreased for each phase of the short-term climate variability covariate, with extreme quantiles increasing by $2$--$4^\circ$C, over different phases. This type of deconstruction/attribution of trends, climate signals into anthropogenic/thermodynamic drivers is itself an important area of climate science. For example, \citet{Fereday2018} study such behaviour for precipitation in typical events using climate model data. There is also a growing literature in extreme event modelling, with a recent review by \cite{smith2025}. 

{Unlike most other studies of climate change, which focus on marginal changes, here we have also focused on making inferences for the changes in spatial risk over time, combining both marginal and dependence features. We found that spatial cold extreme events are becoming less frequent and warmer over time, with this change increasing at more extreme temperatures, and is greatest for low phases of the short-term climatic variability. We have been able to quantify the change in the frequency of spatial extreme events. For example, for the phase of short-term climatic variability (which is most likely to result in such cold temperatures), we estimate that the rate of some stations in Ireland experiencing a colder temperature than $-19.4^\circ$C, the coldest recorded temperature in Ireland, has changed from a 1 in 38-year event in 1950 to a 1 in 3,700-year event in 2022. }

As with any substantive data analysis, there are always plenty of options to take the modelling further. In particular, we see options to do this in terms of capturing seasonality, marginal covariate modelling, evidence that a more advanced type of spatial extreme value model than currently exists may be required to better model the data, that aspects of our $r$-Pareto modelling could be evolved, and that the bootstrap procedure we proposed could be studied systematically. These are considered below.

Our focus has been on extremely cold temperatures in Ireland over the winter period, corresponding to December-February inclusive. For this period, we were able to make a realistic assumption that after accounting for the covariates we have selected, the behaviour of daily minimum temperatures could be modelled as stationarity across these three months across-and-within each year. However, 12\% of the days when the site-wise daily minimum temperature is in the coldest 1\% of all temperatures occur in March, i.e., outside our selection of three months to represent winter. We excluded March data from our analysis as there was clear evidence that extreme cold temperatures in March follow a different marginal behaviour than for the other winter months, even after accounting for covariates. So, to capture the entire period of cold temperature extreme events requires a model evolution from ours, which is capable of incorporating additional seasonality features relative to the monthly HadCRUT5 covariates that we use. We leave this task for future research.        

In all the marginal tail fits throughout the paper, the short-term climatic variability covariate has a coefficient that is constant over space, for all the different forms of short-term climatic variability covariate that we considered. Given that the spatial scale of the covariate covers the whole of Ireland, we anticipated that it should have a common effect on temperature over each station in Ireland. However, to avoid making this assumption,  we could have explored the validity of this assumption by using methods such as \cite{Gelfand2003} for this. It would be interesting to see if this makes any practical difference to key findings in this paper for any sub-region of Ireland.

Our exploratory analysis in Section~\ref{sec:ADvAI} is suggestive that the underlying temperature process may be more complex than any spatial extreme value model that currently exists. Specifically, it appears from using standard diagnostic measures \citep{Coles1999} that asymptotic dependence is applicable up to some dimension, but asymptotic independence seems to be a better description for higher order dimensions. However, this is not feasible for a spatial process, even if it is non-stationary in space. Instead, we appear to be encountering features that are the artefact of the biases in historical sampling locations, with the longest station records coming from coastal sites, and that coastal sites have a much weaker dependence with other sites than simple inter-site distance will account for. Hence, we require the use of either a deformation of the space around coastlines, as in \cite{Richards2021}, or the use of a coastal distance covariate in the variogram of the log-Gaussian processes in the $r$-Pareto process. Nonetheless, having a process which exhibits a mixture of extremal dependence over different spatial scales has close analogies to sparsity in multivariate extremes \citep{Goix2017, Simpson2020, Engelke2021}. A spatial version of that type of mixture modelling approach may also be interesting for data with spatial dependence properties similar to those of our Ireland cold winter temperatures.

A range of more specific spatial dependence features in the paper could also be further explored. In Section~\ref{sec:c6_rparp} we presented our fit of the Mátern variogram assuming the model is invariant to covariates. In fact, we tested for non-stationarity in the scale and range parameters, $\alpha$ and $\phi$, of the Mátern variogram with respect to year, but did not find any statistical evidence for a change. We can see value in trialling the use of the coastal covariate $C(\boldsymbol{s})$ and a short-term climatic variability covariate in the Mátern variogram in future work to explore for departures from each of both spatial and temporal stationarity of the $r$-Pareto process. In Section~\ref{c6:res_spatial_res} we presented estimates of the expected proportion of Ireland with temperatures below $T^{\circ}C$. However, spatial dependence only affects the confidence intervals for these estimates. What would better illustrate, and exploit, our estimated $r$-Pareto process spatial dependence structure would be to estimate this proportion conditionally on a specific site of interest, e.g., Dublin, having temperatures colder than $T^{\circ}C$.

{In Section~\ref{sec:c6_results}, the results are presented conditionally on the short-term climatic variability covariate value for each of the three selected phases. For broader practical use, it would be helpful to integrate out the effect of this covariate for any given time, e.g., as has been performed in univariate extremes by \cite{Eastoe2009}. For our context, if we are interested in a spatial event at time $t$, denoted by $B(t):=\{X_o^{P}(t,\boldsymbol{s})\in B(t,\boldsymbol{s}): \boldsymbol{s}\in \mathcal{S}_o\}$ for some event $B(t,\boldsymbol{s})$ which is extreme for $X_o^{P}(t,\boldsymbol{s})$ for at least some $\boldsymbol{s}\in \mathcal{S}_o$. Then, in particular, in this notation, in our paper, we have presented inferences for}
\[
\Pr\{B(t);h(\tau)\}:= \Pr\{X(t,\boldsymbol{s})\in B(t,\boldsymbol{s}): \boldsymbol{s}\in \mathcal{S}_o \mid
H_{R, m}^I(t)=h(\tau)\}
\]
{for three values of $h(\tau)$ being $\{h_{R}(\tau): \tau=0.1,0.5,0.9\}$ corresponding to the three illustrating phases of the short-term climatic variability. In future work, it might be helpful to estimate unconditional inferences on  $\Pr\{B(t)\}$, with}
\[
\Pr\{B(t)\}=\int_{0}^1\Pr\{B(t);h(\tau)\} \,\mathrm{d}\tau,
\]
{which is not conditional on the phase, or any specific value of $H_{R, m}^I(t)$ of the short-term climatic variability, and with this inference not requiring any further extreme value modelling. }

Although our proposed spatio-temporal bootstrap method appears to be effective from a practical perspective in our setting, it is not standard in form due to it having to address the complex nature of long- and short-range temporal extreme events over space. Despite being confident that it is reasonable to assume that our missing data are missing at random, there are particular issues in bootstrapping due to the high proportion of missing data and with that proportion varying substantially over time. We appreciate that our proposed approach lacks a formal theoretical justification. It might be helpful to consider a simulation-based study to evaluate the robustness, and coverage properties, of this bootstrap method beyond this specific case study, and in that study to assess the comparative performance for different missing data patterns. As such a study would requires considerable extra work for it to be informative more broadly than for our data, we view this study as an interesting area for further work.

\section*{Acknowledgments}
Healy’s work was supported by SFI grant 18/CRT/6049, with some of the work undertaken while in Department of Environmental Sciences, Informatics and Statistics, University of Venice, Ca’ Foscari, Venice, Italia. This study was also carried out within the RISE project and received funding from the European Union Next-GenerationEU - National Recovery and Resilience Plan (NRRP) – MISSION 4 COMPONENT 2, INVESTIMENT 1.1 Fondo per il Programma Nazionale di Ricerca e Progetti di Rilevante Interesse Nazionale (PRIN) – CUP N.H53D23002010006. Andrew Parnell’s work was supported by: the UCD-Met Éireann Research Professorship Programme 28-UCDNWPAI; a Research Ireland Research Centre award 12/RC/2289\_P2; the Research Ireland Centre for Research Training 18CRT/6049; Research Ireland Co-Centre Climate+ in Climate Biodiversity and Water award 22/CC/11103; and research grant number 25/I4I-TC/13542. We thank the editor and referees for their thoughtful comments and suggestions, which greatly helped improve the presentation of this work. We thank Simon Noone (Maynooth University) for help with data. For access to climate data, we acknowledge the World Climate Research Programme’s Working Groups on Regional Climate and on Coupled Modelling, and the European Network for Earth System Modelling. 

\section*{Data Availability}
The observational weather data underlying this article were provided by Met Éireann (for data in the ROI) and the Met Office (for data in NI) under licence. Data will be shared on request to the corresponding author with permission of Met Éireann and the Met Office.

\bibliography{refs}%

@article{HealyProsdocimi2025,
author = {Healy, Dáire and Prosdocimi, Ilaria and Antoniano-Villalobos, Isadora},
title = {{Non-stationarities in extreme hourly precipitation over the Piave basin, northern Italy}},
journal = {Environmetrics},
volume = {36},
number = {8},
pages = {e70051},
year = {2025}
}

@inbook{Smith2025,
   author = {Richard L. Smith},
   city = {London},
   editor = {Miguel de Carvalho and Raphael G. Huser and Philippe Naveau and Brian J. Reich},
   booktitle = {Handbook on Statistics of Extremes},
   publisher = {Chapman \& Hall / CRC series},
   title = {Detection and Attribution of Extreme Weather Events: A Statistical Review},
   year = {2025}
}

@book{little2019statistical,
  title={Statistical {A}nalysis with {M}issing {D}ata},
  author={Little, Roderick JA and Rubin, Donald B},
  volume={793},
  year={2019},
  publisher={John Wiley \& Sons}
}

@article{Cuba2025preprint,
   author = {M. Daniela Cuba and Craig Wilkie and Marian Scott and Daniela Castro-Camilo},
   journal = {arXiv preprint arXiv:2504.20268v1},
   month = {4},
   title = {Spatio-temporal data fusion of censored threshold exceedances},
   year = {2025}
}

@article{Koh2025,
    author = {Koh, Jonathan},
    title = {Contribution to the {Discussion} of ‘{Inference} for extreme spatial temperature events in a changing climate with application to {Ireland}’ by {Healy} et al.},
    journal = {Journal of the Royal Statistical Society Series C: Applied Statistics},
    volume = {74},
    number = {2},
    pages = {315-316},
    year = {2025}
}

@article{castillo-mateo2025,
    author = {Castillo-Mateo, Jorge and Gelfand, Alan E and Cebrián, Ana C and Asín, Jesús},
    title = {Contribution to the {Discussion} of ‘{Inference} for extreme spatial temperature events in a changing climate with application to {Ireland}’ by {Healy} et al.},
    journal = {Journal of the Royal Statistical Society: Series C (Applied Statistics)},
    volume = {74},
    number = {2},
    pages = {309-310},
    year = {2025},
    month = {12}
}

@article{Richards2025,
    author = {Richards, Jordan and Lee, Myung Won and Carcaiso, Viviana and de Carvalho, Miguel},
    title = {Contribution to the {Discussion} of ‘{Inference} for extreme spatial temperature events in a changing climate with application to {Ireland}’ by {Healy} et al.},
    journal = {Journal of the Royal Statistical Society: Series C (Applied Statistics)},
    volume = {74},
    number = {2},
    pages = {307-309},
    year = {2025},
    month = {12}
}

@article {Fereday2018,
      author = "David Fereday and Robin Chadwick and Jeff Knight and Adam A. Scaife",
      title = "Atmospheric Dynamics is the Largest Source of Uncertainty in Future Winter European Rainfall",
      journal = "Journal of Climate",
      year = "2018",
      publisher = "American Meteorological Society",
      address = "Boston MA, USA",
      volume = "31",
      number = "3",
      pages=      "963 - 977",
}

@article{Cuba2025,
    author = {Cuba, Miriam D and Castro-Camilo, Daniela and Scott, Marian E},
    title = {Contribution to the {D}iscussion of ‘{I}nference for extreme spatial temperature events in a changing climate with application to {I}reland’ by {H}ealy et al.},
    journal = {Journal of the Royal Statistical Society: Series C (Applied Statistics)},
    volume = {74},
    number = {2},
    pages = {312-313},
    year = {2025}
}

@article{Huser2025Discuss,
    author = {Huser, Raphaël},
    title = {Seconder of the vote of thanks to {H}ealy et al. and contribution to the {D}iscussion of ‘{I}nference for extreme spatial temperature events in a changing climate with application to {I}reland’},
    journal = {Journal of the Royal Statistical Society: Series C (Applied Statistics)},
    volume = {74},
    number = {2},
    pages = {302-305},
    year = {2025}
}

@article{Simpson2020,
    author = {Simpson, E S and Wadsworth, J L and Tawn, J A},
    title = {Determining the dependence structure of multivariate extremes},
    journal = {Biometrika},
    volume = {107},
    number = {3},
    pages = {513-532},
    year = {2020},
}

@article{Goix2017,
title = {Sparse representation of multivariate extremes with applications to anomaly detection},
journal = {Journal of Multivariate Analysis},
volume = {161},
pages = {12-31},
year = {2017},
author = {Nicolas Goix and Anne Sabourin and Stephan Clémençon}
}

@article{Engelke2021,
   author = "Engelke, Sebastian and Ivanovs, Jevgenijs",
   title = "Sparse Structures for Multivariate Extremes", 
   journal= "Annual Review of Statistics and Its Application",
   year = "2021",
   volume = "8",
   pages = "241-270",
   publisher = "Annual Reviews",
   type = "Journal Article",
  }

@article{Bewentaore2022,
   author = {Sawadogo Béwentaoré and Diakarya Barro},
   number = {1},
   journal = {International Journal of Mathematics and Mathematical Sciences},
   pages = {2608270},
   publisher = {John Wiley \& Sons, Ltd},
   title = {Space-time trend detection and dependence modeling in extreme event approaches by functional peaks-over-thresholds: application to precipitation in {Burkina Faso}},
   volume = {2022},
   year = {2022}
}

@article{Gelfand2003,
author = {Alan E Gelfand and Hyon-Jung Kim and C. F Sirmans and Sudipto Banerjee},
title = {Spatial Modeling With Spatially Varying Coefficient Processes},
journal = {Journal of the American Statistical Association},
volume = {98},
number = {462},
pages = {387--396},
year = {2003},
publisher = {ASA Website},}

@article{Berrocal2010,
   author = {Veronica J. Berrocal and Alan E. Gelfand and David M. Holland},
   number = {2},
   journal = {Journal of Agricultural, Biological, and Environmental Statistics},
   pages = {176-197},
   title = {A spatio-temporal downscaler for output from numerical models},
   volume = {15},
   year = {2010}
}

@article{Huser2025, 
    title={Modeling of spatial extremes in environmental data science: time to move away from max-stable processes}, 
    volume={4},
    journal={Environmental Data Science}, 
    author={Huser, Raphaël and Opitz, Thomas and Wadsworth, Jennifer L.}, year={2025}, 
    pages={e3}
}

@article{Perkins2020,
   author = {S. E. Perkins-Kirkpatrick and S. C. Lewis},

   
   number = {1},
   journal = {Nature Communications},
   pages = {3357},
   pmid = {32620857},
   publisher = {Nature Research},
   title = {Increasing trends in regional heatwaves},
   volume = {11},
   year = {2020}
}

@article{Krock2022,
   author = {Mitchell Krock and Julie Bessac and Michael L. Stein and Adam H. Monahan},

   
   journal = {Weather and Climate Extremes},
   pages = {100438},
   publisher = {Elsevier B.V.},
   title = {Nonstationary seasonal model for daily mean temperature distribution bridging bulk and tails},
   volume = {36},
   year = {2022}
}

@article{Rhines2017,
   author = {Andrew Rhines and Karen A Mckinnon and Martin P Tingley and Peter Huybers},

   number = {3},
   journal = {Journal of Climate},
   pages = {1139-1157},
  title={{Seasonally resolved distributional trends of North American temperatures show contraction of winter variability}},  
  volume = {30},
   year = {2017}
}

@article{Huang2016,
   author = {Whitney K. Huang and Michael L. Stein and David J. McInerney and Shanshan Sun and Elisabeth J. Moyer},
   number = {1},
   journal = {Advances in Statistical Climatology, Meteorology and Oceanography},
   pages = {79-103},
   publisher = {Copernicus GmbH},
   title = {Estimating changes in temperature extremes from millennial-scale climate simulations using generalized extreme value ({GEV}) distributions},
   volume = {2},
   year = {2016}
}

@article{sfpackage,
   author = {Edzer Pebesma},

   
   number = {1},
   journal = {The R Journal},
   pages = {439-446},
   publisher = {Obe and Hsu},
   title = {{Simple features for R: standardized support for spatial vector data}},
   volume = {10},
   year = {2018}
}

@article{Bopp2021,
    title = {{A hierarchical max-infinitely divisible spatial model for extreme precipitation}},
    year = {2021},
    journal = {Journal of the American Statistical Association},
    author = {Bopp, Gregory P. and Shaby, Benjamin A. and Huser, Raphaël},
    number = {533},
    pages = {93--106},
    volume = {116},
    publisher = {Taylor {\&} Francis},

    
    arxivId = {1805.06084}
}

@article{Boucek2016,
    title = {{A review of subtropical community resistance and resilience to extreme cold spells}},
    year = {2016},
    journal = {Ecosphere},
    author = {Boucek, R. E. and Gaiser, E. E. and Liu, H. and Rehage, J. S.},
    number = {10},
    volume = {7},
    publisher = {Ecological Society of America}
}

@article{Screen2014,
    title = {{Amplified mid-latitude planetary waves favour particular regional weather extremes}},
    year = {2014},
    journal = {Nature Climate Change 2014 4:8},
    author = {Screen, James A. and Simmonds, Ian},
    number = {8},
    pages = {704--709},
    volume = {4},
    publisher = {Nature Publishing Group}
}

@article{Kennedy2019,
    title = {{An ensemble data set of sea surface temperature change from 1850: The Met Office Hadley Centre HadSST.4.0.0.0 data set}},
    year = {2019},
    journal = {Journal of Geophysical Research: Atmospheres},
    author = {Kennedy, J. J. and Rayner, N. A. and Atkinson, C. P. and Killick, R. E.},
    number = {14},
    pages = {7719--7763},
    volume = {124}
}

@book{Coles2001,
    title = {{An Introduction to Statistical Modeling of Extreme Values}},
    year = {2001},
    author = {Coles, S. G.},
    edition = {},
    publisher = {Springer},
    address = {London}
}

@article{Morice2021,
    title = {An Updated Assessment of Near-Surface Temperature Change From 1850: The {HadCRUT5} dataset},
    year = {2021},
    journal = {Journal of Geophysical Research: Atmospheres},
    author = {Morice, C. P. and Kennedy, J. J. and Rayner, N. A. and Winn, J. P. and Hogan, E. and Killick, R. E. and Dunn, R. J.H. and Osborn, T. J. and Jones, P. D. and Simpson, I. R.},
    number = {3},
    pages = {e2019JD032361},
    volume = {126},

    
}

@article{McElwain2003,
    title = {{Climate change in Ireland‐ recent trends in temperature and precipitation}},
    year = {2003},
    journal = {Irish Geography},
    author = {McElwain, Laura and Sweeney, John},
    number = {2},
    pages = {97--111},
    volume = {36}
}

@techreport{Garcia2022,
    title = {{Climate Status Report for Ireland 2020}},
    year = {2022},
    author = {Garc{\'{i}}a, Cámaro and C.A, Walther and Dwyer, Ned and Gault, Jeremy},
    institution = {Environmental Protection Agency },
    address = {Dublin},
    isbn = {9781800090095}
}

@article{VanOldenborgh2019,
    title = {{Cold waves are getting milder in the northern midlatitudes}},
    year = {2019},
    journal = {Environmental Research Letters},
    author = {Van Oldenborgh, Geert Jan and Mitchell-Larson, Eli and Vecchi, Gabriel A. and De Vries, Hylke and Vautard, Robert and Otto, Friederike},
    number = {11},
    volume = {14},
    publisher = {Institute of Physics Publishing}
}

@article{Simpson2021b,
    title = {{Conditional modelling of spatio-temporal extremes for Red Sea surface temperatures}},
    year = {2021},
    journal = {Spatial Statistics},
    author = {Simpson, Emma S. and Wadsworth, Jennifer L.},
    pages = {100482},
    volume = {41},
    publisher = {Elsevier B.V.}
}

@article{Donat2014,
    title = {{Consistency of temperature and precipitation extremes across various global gridded in situ and reanalysis datasets}},
    year = {2014},
    journal = {Journal of Climate},
    author = {Donat, Markus G. and Sillmann, Jana and Wild, Simon and Alexander, Lisa V. and Lippmann, Tanya and Zwiers, Francis W.},
    number = {13},
    pages = {5019--5035},
    volume = {27}
}

@misc{CORDEX2019,
    title = {{CORDEX regional climate model data on single levels}},
    year = {2019},
    booktitle = {Copernicus Climate Change Service Climate Data Store},
    author = {{Copernicus Climate Change Service}},
    url = {https://cds.climate.copernicus.eu}
}

@article{Sippel2024,
    title = {{Could an extremely cold central European winter such as 1963 happen again despite climate change?}},
    year = {2024},
    journal = {Weather and Climate Dynamics},
    author = {Sippel, Sebastian and Barnes, Clair and Cadiou, Camille and Fischer, Erich and Kew, Sarah and Kretschmer, Marlene and Philip, Sjoukje and Shepherd, Theodore G. and Singh, Jitendra and Vautard, Robert and Yiou, Pascal},
    number = {3},
    pages = {943--957},
    volume = {5},
    publisher = {Copernicus Publications},
    
}

@article{Coles1999,
    title = {{Dependence measures for extreme value analyses}},
    year = {1999},
    journal = {Extremes},
    author = {Coles, S. G. and Heffernan, Janet E. and Tawn, Jonathan A.},
    number = {4},
    pages = {339--365},
    volume = {2}
}

@article{Wadsworth2012,
    title = {{Dependence modelling for spatial extremes}},
    year = {2012},
    journal = {Biometrika},
    author = {Wadsworth, Jennifer L. and Tawn, Jonathan A.},
    number = {2},
    pages = {253--272},
    volume = {99}
}

@article{dunn2020,
    title = {{Development of an updated global land in situ‐based data set of temperature and precipitation extremes: HadEX3}},
    year = {2020},
    journal = {Journal of Geophysical Research: Atmospheres},
    author = {Dunn, Robert J. H. and Alexander, Lisa V. and Donat, Markus G. and Zhang, Xuebin and Bador, Margot and Herold, Nicholas and Lippmann, Tanya and Allan, Rob and Aguilar, Enric and Barry, Abdoul Aziz and Brunet, Manola and Caesar, John and Chagnaud, Guillaume and Cheng, Vincent and Cinco, Thelma and Durre, Imke and Guzman, Rosaline and Htay, Tin Mar and Wan Ibadullah, Wan Maisarah and Bin Ibrahim, Muhammad Khairul Izzat and Khoshkam, Mahbobeh and Kruger, Andries and Kubota, Hisayuki and Leng, Tan Wee and Lim, Gerald and Li‐Sha, Lim and Marengo, Jose and Mbatha, Sifiso and McGree, Simon and Menne, Matthew and Milagros Skansi, Maria and Ngwenya, Sandile and Nkrumah, Francis and Oonariya, Chalump and Pabon‐Caicedo, Jose Daniel and Panthou, Gérémy and Pham, Cham and Rahimzadeh, Fatemeh and Ramos, Andrea and Salgado, Ernesto and Salinger, Jim and San{\'{e}}, Youssouph and Sopaheluwakan, Ardhasena and Srivastava, Arvind and Sun, Ying and Timbal, Bertrand and Trachow, Nichanun and Trewin, Blair and Schrier, Gerard and Vazquez‐Aguirre, Jorge and Vasquez, Ricardo and Villarroel, Claudia and Vincent, Lucie and Vischel, Theo and Vose, Russ and Bin Hj Yussof, Mohd Noor'Arifin},
    number = {16},
    month = {8},
    pages = {e2019JD032263},
    volume = {125},
    publisher = {Blackwell Publishing Ltd}
}

@article{Hall2015,
    title = {{Drivers of North Atlantic polar front jet stream variability}},
    year = {2015},
    journal = {International Journal of Climatology},
    author = {Hall, Richard and Erd{\'{e}}lyi, Róbert and Hanna, Edward and Jones, Julie M. and Scaife, Adam A.},
    number = {8},
    pages = {1697--1720},
    volume = {35},
    publisher = {John Wiley {\&} Sons, Ltd}
}

@article{Vihma2020,
    title = {{Effects of the tropospheric large-scale circulation on European winter temperatures during the period of amplified Arctic warming}},
    year = {2020},
    journal = {International Journal of Climatology},
    author = {Vihma, Timo and Graversen, Rune and Chen, Linling and Handorf, Dörthe and Skific, Natasa and Francis, Jennifer A. and Tyrrell, Nicholas and Hall, Richard and Hanna, Edward and Uotila, Petteri and Dethloff, Klaus and Karpechko, Alexey Y. and Bj{\"{o}}rnsson, Halldor and Overland, James E.},
    number = {1},
    pages = {509--529},
    volume = {40},
    publisher = {John Wiley and Sons Ltd}
}

@article{sousa2018,
    title = {{European temperature responses to blocking and ridge regional patterns}},
    year = {2018},
    journal = {Climate Dynamics},
    author = {Sousa, Pedro M and Trigo, Ricardo M and Barriopedro, David and Soares, Pedro MM and Santos, Jo{\~a}o A},
    pages = {457--477},
    volume = {50}
}

@article{hooker2008,
    title = {{Evaluation of cover crop and reduced cultivation for reducing nitrate leaching in Ireland}},
    year = {2008},
    journal = {Journal of Environmental Quality},
    author = {Hooker, K. V. and Coxon, C. E. and Hackett, R. and Kirwan, L. E. and O'Keeffe, E. and Richards, K.G.},
    number = {1},
    month = {1},
    pages = {138--145},
    volume = {37},
    publisher = {Wiley},

    
    pmid = {18178886}
}

@article{Francis2012,
    title = {{Evidence linking Arctic amplification to extreme weather in mid-latitudes}},
    year = {2012},
    journal = {Geophysical Research Letters},
    author = {Francis, Jennifer A. and Vavrus, Stephen J.},
    number = {6},
    month = {3},
    volume = {39},
    publisher = {John Wiley {\&} Sons, Ltd}
}

@book{Hoskins2014,
    title = {{Fluid Dynamics of the Midlatitude Atmosphere}},
    year = {2014},
    author = {Hoskins, Brian J. and James, Ian N.},
    publisher = {John Wiley {\&} Sons, Ltd}
}

@article{DeFondeville2022,
    title = {{Functional peaks-over-threshold analysis}},
    year = {2022},
    journal = {Journal of the Royal Statistical Society: Series B (Statistical Methodology)},
    author = {de Fondeville, Raphaël and Davison, Anthony C.},
    number = {4},
    pages = {1392--1422},
    volume = {84},
    arxivId = {2002.02711}
}

@article{Dombry2015,
  title={Functional regular variations, {{Pareto}} processes and peaks over threshold},
  author={Dombry, Cl{\'e}ment and Ribatet, Mathieu},
  journal={Statistics and its Interface},
  volume={8},
  number={1},
  pages={9--17},
  year={2015},
  publisher={International Press of Boston}
}

@article{Chavez-Demoulin2005,
    title = {{Generalized additive modelling of sample extremes}},
    year = {2005},
    journal = {Journal of the Royal Statistical Society: Series C (Applied Statistics)},
    author = {Chavez-Demoulin, V. and Davison, A. C.},
    number = {1},
    pages = {207--222},
    volume = {54}
}

@article{Youngman2019,
    title = {{Generalized additive models for exceedances of high thresholds with an application to return level estimation for U.S. wind gusts}},
    year = {2019},
    journal = {Journal of the American Statistical Association},
    author = {Youngman, Benjamin D.},
    number = {528},
    pages = {1865--1879},
    volume = {114},
    publisher = {Taylor {\&} Francis},
}

@article{Palacios2020,
    title = {{Generalized {Pareto} processes for simulating space-time extreme events: an application to precipitation reanalyses}},
    year = {2020},
    journal = {Stochastic Environmental Research and Risk Assessment},
    author = {Palacios-Rodr{\'{i}}guez, F. and Toulemonde, G. and Carreau, J. and Opitz, T.},
    number = {12},
    month = {12},
    pages = {2033--2052},
    volume = {34},
    publisher = {Springer Science and Business Media Deutschland GmbH}
}

@article{Ballester2023,
    title = {{Heat-related mortality in Europe during the summer of 2022}},
    year = {2023},
    journal = {Nature Medicine},
    author = {Ballester, Joan and Quijal-Zamorano, Marcos and M{\'{e}}ndez Turrubiates, Raúl Fernando and Pegenaute, Ferran and Herrmann, François R. and Robine, Jean Marie and Basaga{\~{n}}a, Xavier and Tonne, Cathryn and Ant{\'{o}}, Josep M. and Achebak, Hicham},
    number = {7},
    month = {7},
    pages = {1857--1866},
    volume = {29},

    
}

@book{banerjee2014hierarchical,
    title = {{Hierarchical Modeling and Analysis for Spatial Data}},
    year = {2014},
    author = {Banerjee, Sudipto and Carlin, Bradley P and Gelfand, Alan E},
    edition = {},

    publisher = {Chapman {\&} Hall/CRC},
    address = {New York}
}

@article{DeFondeville2018,
    title = {{High-dimensional peaks-over-threshold inference}},
    year = {2018},
    journal = {Biometrika},
    author = {de Fondeville, R. and Davison, A. C.},
    number = {3},
    pages = {575--592},
    volume = {105},

    
    arxivId = {1605.08558}
}

@techreport{Nolan2020,
    title = {{High-resolution Climate Projections for Ireland-A Multi-model Ensemble Approach}},
    year = {2020},
    author = {Nolan, Paul and Flanagan, Jason},
    institution = {EPA},
    address = {Dublin}
}

@article{Wadsworth2019,
    title = {{Higher-dimensional spatial extremes via single site conditioning}},
    year = {2022},
    journal = {Spatial Statistics},
    author = {Wadsworth, Jennifer L. and Tawn, Jonathan A.},
    number = {100677},
    volume = {51},
    arxivId = {1912.06560}
}

@article{Healy2023,
   title = {Inference for extreme spatial temperature events in a changing climate with application to {Ireland} (with discussion)},
   year = {2025},
    author = {Healy, Dáire and Tawn, Jonathan and Thorne, Peter and Parnell, Andrew},
   number = {2},
   journal = {Journal of the Royal Statistical Society: Series C (Applied Statistics)},
   pages = {275-299},
   volume = {74}
}

@techreport{Mateus2022,
    title = {{Isotherman maps of maximum and minimum shade air temperatures in Ireland}},
    year = {2022},
    author = {Mateus, Carla and Coonan, Barry},
    institution = {Met {\'{E}}ireann},
    address = {Dublin}
}

@article{Osborn2021,
    title = {{Land surface air temperature variations across the globe updated to 2019: the CRUTEM5 data set}},
    year = {2021},
    journal = {Journal of Geophysical Research: Atmospheres},
    author = {Osborn, T. J. and Jones, P. D. and Lister, D. H. and Morice, C. P. and Simpson, I. R. and Winn, J. P. and Hogan, E. and Harris, I. C.},
    number = {2},
    pages = {e2019JD032352},
    volume = {126}
}

@article{Huser2021,
    title = {{Max‐infinitely divisible models and inference for spatial extremes}},
    year = {2021},
    journal = {Scandinavian Journal of Statistics},
    author = {Huser, Raphael and Opitz, Thomas and Thibaud, Emeric},
    number = {1},
    month = {1},
    pages = {321--348},
    volume = {48},
    arxivId = {1801.02946}
}

@article{ukmo-midas,
    title = {{Met Office Integrated Data Archive System (MIDAS) land and marine surface stations data (1853-current)}},
    year = {2012},
    journal = {NCAS British Atmospheric Data Centre},
    author = {{Met Office}},
    url = {{https://catalogue.ceda.ac.uk/uuid/916ac4bbc46f7685ae9a5e10451bae7c}}
}

@article{Eastoe2009,
    title = {{Modelling non-stationary extremes with application to surface level ozone}},
    year = {2009},
    journal = {Journal of the Royal Statistical Society: Series C (Applied Statistics)},
    author = {Eastoe, Emma F. and Tawn, Jonathan A.},
    number = {1},
    pages = {25--45},
    volume = {58}
}

@article{Davison1990,
    title = {{Models for exceedances over high thresholds (with discussion)}},
    year = {1990},
    journal = {Journal of the Royal Statistical Society: Series B (Statistical Methodology)},
    author = {Davison, A. C. and Smith, R. L.},
    number = {3},
    pages = {393--442},
    volume = {52},
}

@article{DeFondeville2021,
  title={mvPot: Multivariate peaks-over-threshold modelling for spatial extreme events},
  author={de Fondeville, Raphael and Belzile, Leo and Thibaud, E},
  journal={R package},
  year={2021}
}

@article{ryan2017,
    title = {{On some Mat{\'{e}}rn covariance functions for spatio-temporal random fields}},
    year = {2017},
    journal = {Statistica Sinica},
    author = {Ip, Ryan H. L. and Li, W. K.},
    number = {2},

    pages = {805--822},
    volume = {27}
}

@article{Buishand2008,
    title = {{On spatial extremes: with application to a rainfall problem}},
    year = {2008},
    journal = {The Annals of Applied Statistics},
    author = {Buishand, T. A. and de Haan, L. and Zhou, C.},
    number = {2},
    month = {6},
    pages = {624--642},
    volume = {2},
    publisher = {Institute of Mathematical Statistics}
}

@article{Deser2000,
    title = {{On the teleconnectivity of the Arctic Oscillation}},
    year = {2000},
    journal = {Geophysical Research Letters},
    author = {Deser, C.},
    number = {6},
    pages = {779--782},
    volume = {27},
    publisher = {American Geophysical Union},

    
}

@article{Gerber2009,
    title = {{On the zonal structure of the North Atlantic oscillation and annular modes}},
    year = {2009},
    journal = {Journal of the Atmospheric Sciences},
    author = {Gerber, Edwin P. and Vallis, Geoffrey K.},
    number = {2},
    pages = {332--352},
    volume = {66},

    
}

@article{Conlon2011,
  title={Preventing cold-related morbidity and mortality in a changing climate},
  author={Conlon, Kathryn C and Rajkovich, Nicholas B and White-Newsome, Jalonne L and Larsen, Larissa and O’Neill, Marie S},
  journal={Maturitas},
  volume={69},
  number={3},
  pages={197--202},
  year={2011},
  publisher={Elsevier}
}

@article{Heidrun2015,
    title = {{Recent changes in Arctic temperature extremes: warm and coldspells during winter and summer}},
    year = {2015},
    journal = {Environmental Research Letters},
    author = {Matthes, Heidrun and Rinke, Annette and Dethloff, Klaus},
    number = {},
    pages = {114020},
    volume = {10},
    publisher = {Institute of Physics Publishing},
    
}

@article{Thompson2021,
    title = {{Regional climate impacts of the Northern Hemisphere annular mode}},
    year = {2001},
    journal = {Science},
    author = {Thompson, D. W.J. and Wallace, J. M.},
    number = {5527},
    pages = {85--89},
    volume = {293},
    publisher = {American Association for the Advancement of Science},

    
    pmid = {11441178}
}

@incollection{Liou2002,
title = {{Chapter 2: Solar Radiation at the Top of the Atmosphere}},
author = {K.N. Liou},
editor = {K.N. Liou},
series = {International Geophysics},
publisher = {Academic Press},
volume = {84},
pages = {37-64},
year = {2002},
booktitle = {An Introduction to Atmospheric Radiation},
issn = {0074-6142}}

@article{Richards2021,
    title = {{Spatial deformation for nonstationary extremal dependence}},
    year = {2021},
    journal = {Environmetrics},
    author = {Richards, Jordan and Wadsworth, Jennifer L.},
    number = {5},
    pages = {1--22},
    volume = {32},

    
    arxivId = {2101.07167}
}

@article{Coles1994,
    title = {{Statistical methods for multivariate extremes: an application to structural design (with discussion)}},
    year = {1994},
    journal = {Journal of the Royal Statistical Society: Series C (Applied Statistics)},
    author = {Coles, S. G. and Tawn, Jonathan A.},
    number = {1},
    pages = {1--48},
    volume = {43}
}

@article{He2022,
    title = {{Substantial increase of compound droughts and heatwaves in wheat growing seasons worldwide}},
    year = {2022},
    journal = {International Journal of Climatology},
    author = {He, Yan and Fang, Jiayi and Xu, Wei and Shi, Peijun},
    number = {10},
    month = {8},
    pages = {5038--5054},
    volume = {42},
    publisher = {John Wiley and Sons Ltd}
}

@article{Thompson1998,
    title = {{The Arctic oscillation signature in the wintertime geopotential height and temperature fields}},
    year = {1998},
    journal = {Geophysical Research Letters},
    author = {Thompson, David W.J. and Wallace, John M.},
    number = {9},
    pages = {1297--1300},
    volume = {25},
    publisher = {John Wiley {\&} Sons, Ltd}
}

@book{Rohan1986,
    title = {{The Climate of Ireland}},
    year = {1986},
    author = {Rohan, Patrick K},
    edition = {2},
    volume = {},
    publisher = {Stationery Office},
    address = {Dublin}
}

@article{Ferreira2014,
    title = {{The generalized {Pareto} process; with a view towards application and simulation}},
    year = {2014},
    journal = {Bernoulli},
    author = {Ferreira, Ana and de Haan, Laurens},
    number = {4},
    pages = {1717--1737},
    volume = {20}
}

@article{Hickey2011,
    title = {{The historic record of cold spells in Ireland}},
    year = {2011},
    journal = {Irish Geography},
    author = {Hickey, Kieran},
    number = {2-3},
    pages = {303--321},
    volume = {44}
}

@article{Skend2021,
   author = {Sandra Skendžić and Monika Zovko and Ivana Pajač Živković and Vinko Lešić and Darija Lemić},

   
   number = {5},
   journal = {Insects},
   pages = {440-471},
   publisher = {MDPI AG},
   title = {The impact of climate change on agricultural insect pests},
   volume = {12},
   year = {2021}
}

@incollection{Stendel2021,
    title = {{The jet stream and climate change}},
    year = {2021},
    booktitle = {{Climate Change: Observed Impacts on Planet Earth, Third Edition}},
    editor = {Letcher, Trevor},
    author = {Stendel, Martin and Francis, Jennifer and White, Rachel and Williams, Paul D. and Woollings, Tim},
    pages = {327--357},
    publisher = {Elsevier}
}

@incollection{IPCC2021,
    title = {{The Physical Science Basis. Contribution of Working Group I to the Sixth Assessment Report of the Intergovernmental Panel on Climate Change}},
    year = {2021},
    booktitle = {Climate Change 2021},
    author = {{IPCC}},
    editor = {Masson-Delmotte, V and Zhai, P and Pirani, A and Connors, S.L and P{\'{e}}an, C and Berger, S and Caud, N and Chen, Y and Goldfarb, L and Gomis, M.I and Huang, M and Leitzell, K and Lonnoy, E and Matthews, J.B.R and Maycock, T.K and Waterfield, T and Yelek{\c{c}}i, O and Yu, R and Zhou, B},
    pages = {2061--2086},
    publisher = {Cambridge University Press},
    address = {Cambridge, United Kingdom and New York, NY, USA},
}

@article{bindi2011,
    title = {{The responses of agriculture in Europe to climate change}},
    year = {2011},
    journal = {Regional Environmental Change},
    author = {Bindi, Marco and Olesen, Jørgen E.},
    number = {SUPPL. 1},
    pages = {151--158},
    volume = {11}
}

@article{Giorgi2019,
    title = {{Thirty years of regional climate modeling: where are we and where are we going next?}},
    year = {2019},
    journal = {Journal of Geophysical Research: Atmospheres},
    author = {Giorgi, Filippo},
    number = {11},
    pages = {5696--5723},
    volume = {124}
}

@article{Osland2021,
    title = {{Tropicalization of temperate ecosystems in North America: The northward range expansion of tropical organisms in response to warming winter temperatures}},
    year = {2021},
    journal = {Global Change Biology},
    author = {Osland, Michael J. and Stevens, Philip W. and Lamont, Margaret M. and Brusca, Richard C. and Hart, Kristen M. and Waddle, J. Hardin and Langtimm, Catherine A. and Williams, Caroline M. and Keim, Barry D. and Terando, Adam J. and Reyier, Eric A. and Marshall, Katie E. and Loik, Michael E. and Boucek, Ross E. and Lewis, Amanda B. and Seminoff, Jeffrey A.},
    number = {13},
    pages = {3009--3034},
    volume = {27},
    publisher = {Blackwell Publishing Ltd}
}

@incollection{SenGupta2012,
    title = {{Variability and change in the ocean}},
    year = {2012},
    booktitle = {{The Future of the World's Climate}},
    author = {Sen Gupta, Alex and McNeil, Ben},
    pages = {141--165},
    publisher = {Elsevier},
    editor =  {Henderson-Sellers, Ann and McGuffie, Kendal},
}

@article{CHRISTIANSEN2018,
    title = {{Was the cold European winter of 2009/10 modified by anthropogenic climate change? An attribution study}},
    year = {2018},
    journal = {Journal of Climate},
    author = {Christiansen, B O and Alvarez-Castro, Carmen and Christidis, Nikolaos and Ciavarella, Andrew and Colfescu, Ioana and Cowan, Tim and Eden, Jonathan and Hauser, Mathias and Hempelmann, Nils and Klehmet, Katharina and Lott, Fraser and Nangini, Cathy and Jan Van Oldenborgh, Geert and Orth, René and Stott, Peter and Tett, Simon and Vautard, Robert and Wilcox, Laura and Yiou, Pascal},
    number = {9},
    pages = {3387--3410},
    volume = {31},
}

@article{Cattiaux2010,
    title = {{Winter 2010 in Europe: a cold extreme in a warming climate}},
    year = {2010},
    journal = {Geophysical Research Letters},
    author = {Cattiaux, J. and Vautard, R. and Cassou, C. and Yiou, P. and Masson-Delmotte, V. and Codron, F.},
    number = {20},
    volume = {37},
    publisher = {Blackwell Publishing Ltd},

    
}

@techreport{Meteireann2019,
    title = {{Winter report 2018/19}},
    year = {2019},
    author = {{Met {\'{E}}ireann}},
    institution = {Met {\'{E}}ireann},
    address = {Dublin}
}

\end{document}